\documentclass[aps,prb,reprint,superscriptaddress,floatfix,longbibliography]{revtex4-2}

\usepackage{amsmath,amssymb,bm,graphicx}
\usepackage[usenames,dvipsnames]{xcolor}
\usepackage{hyperref}
\hypersetup{
  colorlinks=true,
  linkcolor=MidnightBlue,
  citecolor=MidnightBlue,
  urlcolor=MidnightBlue
}
\usepackage{times}   % APS recommends Times-like font for PRB

\usepackage{microtype}
\usepackage{physics}
\usepackage{siunitx}
\begin{document}

%---------------------------  Title  -------------------------------
\title{Mobile impurity coupled to correlated lattice bosons}

\author{Chao Zhang}
\email{chaozhang@ahnu.edu.cn}
\affiliation{Department of Physics, Anhui Normal University, Wuhu 241002, China}

%---------------------------  Abstract  ----------------------------
\begin{abstract}
We investigate how the coherence and spatial dressing of a single impurity evolve in the two-dimensional
Bose–Hubbard model when the impurity couples attractively to the bath.
Using large-scale, sign-problem-free worm-algorithm quantum Monte Carlo, we measure the impurity winding,
bath superfluid response and compressibility, and impurity--bath density correlations. In a compressible superfluid bath ($U_{\mathrm{b}}/t=13.3$), strengthening attraction drives an interaction-controlled
\emph{winding-collapse self-trapping crossover}: a mobile light polaron evolves continuously into a heavy polaron and
ultimately into a bound cluster with vanishing winding, while the bath remains globally superfluid.
In incompressible Mott-insulating baths, by contrast, extended density rearrangements are suppressed and the dressing
cloud collapses; we compare the resulting short-range deformation patterns for both attractive and repulsive couplings.
Across the SF--MI transition at fixed moderate attraction ($U_{\mathrm{ib}}/t=-8.0$), the impurity crosses over from a
mobile polaron with an extended deformation cloud to a nearly free defect with minimal dressing in the Mott background;
for this coupling it does not lock into a fully self-trapped defect even deep in the insulator. Together with our companion Letter [\emph{Impurity Self-Trapping in Lattice Bose systems}] on repulsive couplings,
these results provide a unified microscopic picture of impurity self-trapping in correlated lattice bosons, connecting
interaction-driven winding collapse in the superfluid to compressibility-controlled undressing across the SF--MI transition.
\end{abstract}

\maketitle

\section{Introduction}

The behavior of a single impurity immersed in a quantum many-body medium
provides a sensitive probe of correlations, coherence, and emergent quasiparticles.
In weakly interacting Bose gases, such an impurity forms a Bose polaron:
a dressed quasiparticle characterized by a finite quasiparticle residue,
a renormalized effective mass, and an extended bath-density deformation cloud~\cite{Grusdt_2025, ScazzaZaccanti2022, ArdilaGiorgini2015, ArdilaGiorgini2016, Ardila2019, Ardila2020, PhysRevLett.120.050405, PhysRevA.109.013325, PhysRevA.98.063631}.
In the continuum, Bose polarons have been observed and characterized with high precision using
radio-frequency and Ramsey spectroscopy in ultracold atomic mixtures~\cite{JorgensenPRL2016, HuPRL2016, YanScience2020},
and their properties across weak to strong coupling have been mapped out both experimentally and theoretically~\cite{Grusdt_2025}.
These studies have established the continuum Bose polaron as a paradigmatic impurity problem in a quantum fluid.

Beyond continuum Bose gases, optical lattices offer a complementary platform where the bath is naturally described by the
Bose--Hubbard model (BHM)~\cite{JakschPRL1998, PhysRevB.40.546, PhysRevB.75.134302, PhysRevA.77.015602},
providing a controlled setting to study a single mobile impurity in a strongly correlated lattice Bose fluid.
By tuning the ratio between on-site repulsion and tunneling, the BHM realizes a quantum phase transition from a compressible
superfluid (SF) to an incompressible Mott insulator (MI), as first observed in pioneering cold-atom experiments~\cite{GreinerNature2002}.
With the advent of quantum gas microscopy, site-resolved imaging and control of individual atoms in optical lattices has become possible~\cite{BakrScience2010, ShersonNature2010},
including the preparation and tracking of a single impurity or spin excitation in Bose--Hubbard chains and two-dimensional lattices~\cite{FukuharaNatPhys2013}.
This opens the door to directly visualizing how an impurity propagates, dresses, and eventually \emph{self-traps} as the background evolves from SF to MI.

Theoretical work on lattice Bose polarons and impurities in the BHM has addressed a number of important limits~\cite{10.21468/SciPostPhys.19.1.002, Santiago-García_2024, DuttaMueller2013, ColussiPRL2023, KeilerNJP2020, DingSciPost2023}.
Variational approaches have explored mobile impurities near the SF--MI transition~\cite{ColussiPRL2023},
and recent large-scale quantum Monte Carlo (QMC) simulations have analyzed the ``Bose polaron'' from weak to strong impurity--bath coupling in two dimensions~\cite{Hartweg2025}.
In particular, Ref.~\cite{Hartweg2025} used worm-algorithm QMC in the grand-canonical ensemble to show that, at weak coupling,
polaron spectral properties such as the effective mass can serve as a sensitive probe of the MI--SF transition,
while at strong coupling the impurity binds an extra particle or hole in the MI, leading to bound states beyond the simple polaron picture.
Nevertheless, most lattice studies to date have focused primarily on \emph{impurity} spectral observables---energy shifts, effective masses, and spectral functions---and typically emphasized either the SF side or the MI side separately,
with the bath entering mainly through averaged properties.
What is still lacking is a unified, real-space characterization that follows \emph{both} the impurity's coherent mobility and the associated impurity-centered bath-density deformations across the SF--MI transition.

In a recent companion Letter~\cite{chaoletter}, we addressed the fate of a single mobile impurity in the two-dimensional Bose--Hubbard model at unit filling by mapping out its regimes in the plane of bath interaction $U_{\mathrm{b}}/t$ and local impurity--bath coupling $U_{\mathrm{ib}}/t$.
Using sign-problem-free multi-species worm-algorithm QMC simulations in the grand-canonical ensemble~\cite{ProkofevJETP1998, secondworm, Lingua2018},
we showed that impurity \emph{self-trapping} in this setting is organized by two distinct routes.
First, within a compressible SF bath, increasing $|U_{\mathrm{ib}}|/t$ drives an \emph{interaction-driven winding-collapse crossover}:
a weakly coupled impurity forms a mobile light polaron with finite winding and an extended dressing cloud, whereas stronger coupling suppresses winding while the bath remains globally superfluid, producing a heavy, self-trapped state and, for repulsive $U_{\mathrm{ib}}/t>0$, eventually a depletion \emph{saturated-bubble} configuration.
Second, along trajectories where the bath is tuned across the SF--MI transition at fixed $U_{\mathrm{ib}}/t$, the evolution becomes \emph{compressibility controlled}.
Starting from a compressible SF, the impurity appears as a mobile light polaron with an extended deformation cloud, while increasing $U_{\mathrm{b}}/t$ progressively suppresses the bath superfluid response and compressibility and reduces the impurity winding.
Upon entering the MI regime, the polaronic dressing collapses and the impurity becomes nearly free in an incompressible background; deeper in the MI, strong coupling can lock the impurity to a vacancy- or particle-type defect with a strongly localized deformation cloud.
These trends are captured directly by the impurity-centered cumulative bath-density deformation $\Delta N(R)$, which evolves from a broad polaronic profile in the SF to a nearly flat response in the MI and, at strong coupling, to a compact defect signature.
In this way, the Letter established a global $U_{\mathrm{b}}/t$--$U_{\mathrm{ib}}/t$ diagram connecting light polarons, heavy polarons, and saturated-bubble/bound-cluster regimes in the SF, as well as nearly free and defect-bound self-trapped states in the MI.
It further clarified that self-trapping within the SF proceeds via a continuous crossover (without a bath phase transition), whereas in the MI the change between nearly free and defect-bound sectors can become sharp, while the evolution across the SF--MI boundary at fixed $U_{\mathrm{ib}}/t$ remains a smooth, compressibility-controlled crossover.

While the Letter~\cite{chaoletter} establishes a broad global picture of impurity \emph{self-trapping}
in the Bose--Hubbard model, it does not fully develop a microscopic description for attractive impurity--bath couplings,
nor for impurity behavior deep in a genuinely \emph{incompressible} Mott-insulating (MI) bath.
In the MI regime, long-wavelength density fluctuations are gapped and the superfluid stiffness vanishes~\cite{PhysRevA.77.015602}.
At weak $U_{\mathrm{ib}}/t$, an impurity is therefore expected to behave nearly as a free defect on top of a rigid background,
whereas at strong $|U_{\mathrm{ib}}|/t$ it can bind and pin a particle- or hole-like defect, as suggested in Ref.~\cite{Hartweg2025}.
Although the global diagram in Ref.~\cite{chaoletter} already indicated where the interaction-driven winding-collapse route in the SF
and the compressibility-controlled route across the SF--MI boundary meet inside the MI region, several key microscopic questions remained open:
(i)~How does the interaction-driven winding-collapse mechanism manifest for \emph{attractive} impurity--bath couplings in a superfluid background?
(ii)~Once the bath is strictly incompressible, in what ways do attractive and repulsive impurities differ in their deformation clouds and defect-binding patterns?
(iii)~For a fixed attractive coupling, how does the compressibility-controlled \emph{self-trapping} route proceed as the bath is tuned from SF into the MI?

In this paper, we build on Ref.~\cite{chaoletter} and address these questions with a systematic, numerically exact worm-algorithm QMC study
that focuses on attractive impurity--bath couplings and provides a detailed account of impurity behavior in MI baths.
We analyze three complementary trajectories.
(i)~In a compressible SF bath, we implement an \emph{interaction-driven winding-collapse} protocol for attractive impurities by fixing
$U_{\mathrm{b}}/t=13.33$ and tuning $U_{\mathrm{ib}}/t$ from $-1.0$ to $-40.0$, thereby following the evolution from a mobile light polaron with finite winding
to a heavy polaron and ultimately to a bound cluster while the bath remains globally superfluid.
(ii)~Inside the MI regime, we compare attractive and repulsive couplings and show how the impurity crosses over from a nearly free defect
to a defect-bound, self-trapped state (particle- or vacancy-like) as $|U_{\mathrm{ib}}|/t$ increases in an incompressible background.
(iii)~To isolate the role of bath compressibility for an attractive impurity, we fix $U_{\mathrm{ib}}/t=-8.0$ and tune $U_{\mathrm{b}}/t$ from deep SF,
through the critical region, into deep MI, tracking how impurity winding and the deformation cloud collapse as the bath becomes incompressible.
For each parameter set we measure the impurity winding number, the bath superfluid response and compressibility,
and impurity-centered real-space observables including the radial correlator $C_{\rm ib}(R)$ and the cumulative bath-density deformation $\Delta N(R)$.
Together with Ref.~\cite{chaoletter}, these results promote the global two-route picture into a fully microscopic account that spans
both attractive and repulsive couplings and both compressible and incompressible bath regimes.

Our results complement and extend previous work in several ways.
First, by combining worldline winding diagnostics with impurity-centered real-space observables ($C_{\rm ib}(R)$ and $\Delta N(R)$),
we directly visualize the \emph{self-trapping} process in both SF and MI backgrounds, rather than inferring it solely from effective masses or ground-state energies.
Second, we clarify how interaction-driven winding collapse and compressibility-controlled undressing operate microscopically for attractive impurities:
the same winding-based criterion, together with the deformation cloud, tracks the evolution from a mobile light polaron to a heavy polaron
and finally to a bound cluster in the SF, and its subsequent modification as the bath loses compressibility.
Third, by following the impurity across the SF--MI transition and into the deep MI regime, we show how the light polaron in the SF evolves continuously into a nearly free defect
and, at stronger coupling in the MI, into a defect-bound self-trapped state, thereby connecting the polaronic and defect descriptions within a single framework.
Given the rapid experimental progress in site-resolved control of impurities in optical lattices~\cite{BakrScience2010, ShersonNature2010, FukuharaNatPhys2013},
we expect that the observables and regimes identified here can be probed directly in future quantum-gas-microscope experiments.

The remainder of this paper is organized as follows.
Section~\ref{model} introduces the two-component Bose--Hubbard Hamiltonian, and Sec.~\ref{method} describes the QMC method and defines the impurity and bath observables,
including the impurity-centered correlator $C_{\mathrm{ib}}(\mathbf r)$ and the cumulative bath-density deformation $\Delta N(R)$.
Sections~\ref{sec:SF}, \ref{sec:MI}, and \ref{sec:crossing} present our main numerical results.
We first briefly summarize the global impurity diagram established in Ref.~\cite{chaoletter}, and then analyze in detail three trajectories:
(i)~an interaction-driven winding-collapse path in an SF bath with attractive $U_{\mathrm{ib}}/t$,
(ii)~impurity physics in an MI bath for representative attractive and repulsive couplings $U_{\mathrm{ib}}/t=\pm 8.0$,
and (iii)~a compressibility-controlled path where the bath is tuned from SF to MI at fixed $U_{\mathrm{ib}}/t=-8.0$.
A summary of the main conclusions and an outlook are given in Sec.~\ref{outlook}.

\section{Model}
\label{model}

We study a single mobile impurity immersed in a two-dimensional
Bose--Hubbard bath on a square lattice described by the Hamiltonian
\begin{align}
H = & 
-t_{\mathrm{imp}} \!\!\sum_{\langle i,j\rangle}
   (a_i^\dagger a_j + \mathrm{H.c.})
 + U_{\mathrm{ib}}\sum_i n_{\mathrm{imp},i} n_{\mathrm{b},i} \nonumber \\
&  -t_{\mathrm{b}} \!\!\sum_{\langle i,j\rangle}
   (b_i^\dagger b_j + \mathrm{H.c.})
 + \frac{U_{\mathrm{b}}}{2}\sum_i n_{\mathrm{b},i}(n_{\mathrm{b},i}-1) \nonumber \\
& - \mu_{\mathrm{b}}\sum_i n_{\mathrm{b},i} ,
\label{eq:Hamiltonian}
\end{align}
where \(b_i^\dagger\) (\(a_i^\dagger\)) creates a bath (impurity) boson
on lattice site \(i\),
and \(n_{\mathrm{b},i}=b_i^\dagger b_i\),
\(n_{\mathrm{imp},i}=a_i^\dagger a_i\) are the corresponding
on-site number operators.
The bath hopping amplitude \(t_{\mathrm{b}}\) sets the energy scale
(\(t_{\mathrm{b}}=1\) throughout),
and the on-site repulsion \(U_{\mathrm{b}}\) controls the
superfluid–Mott insulator transition of the bath.
The impurity hopping \(t_{\mathrm{imp}}=t=1\) is taken equal to \(t_{\mathrm{b}}\).
The local impurity–bath coupling \(U_{\mathrm{ib}}\)
is the key tuning parameter:
\(U_{\mathrm{ib}}>0\) corresponds to repulsion,
\(U_{\mathrm{ib}}<0\) to attraction.
The bath chemical potential \(\mu_{\mathrm{b}}\) is adjusted
to realize unit filling of the bath, while a single impurity is enforced by
\(\sum_i\langle n_{\mathrm{imp},i}\rangle=1\).

\begin{figure}[t]
    \centering
    \includegraphics[width=\linewidth]{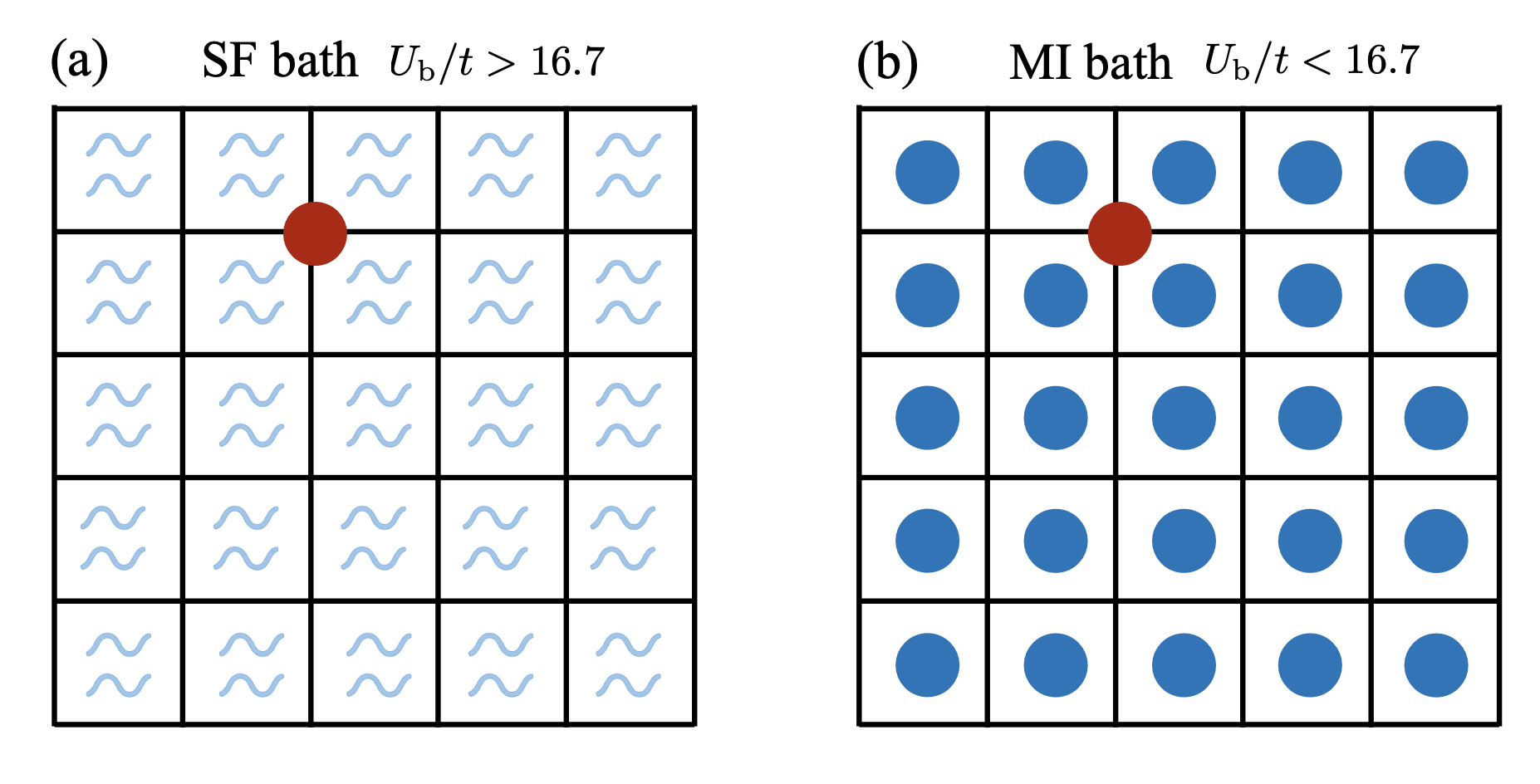}
    \caption{\textbf{Schematic impurity behavior in superfluid (SF) and Mott-insulating (MI) baths.}
    The bath undergoes a SF--MI transition at $U_{\mathrm{b}}/t \simeq 16.7$.
    (a)~In a compressible SF bath, bosons (blue wavy background) possess long-range phase coherence and support extended density fluctuations. 
    A repulsive impurity (red disk) creates a broad depletion cloud, whereas an attractive impurity induces a smooth accumulation halo. 
    Because the bath is compressible, these density modulations spread over many sites and the impurity retains coherent motion with finite winding, forming a light polaron at weak coupling and an extended heavy state at stronger coupling.
    (b)~In an incompressible MI bath, density fluctuations are frozen and only short-range particle--hole excitations are allowed. 
    For both repulsive and attractive impurity--bath couplings, the impurity behaves as a nearly free defect at moderate $U_{\mathrm{b}}/t$, producing only weak local distortion. 
    At larger $U_{\mathrm{b}}/t$ the impurity becomes fully self-trapped: a repulsive impurity expels one bath particle and forms a quantized vacancy defect, while an attractive impurity binds extra one boson into a particle defect. 
    Both MI-side defects exhibit short-range density distortion and vanishing impurity winding, in stark contrast to the extended polaronic dressing in the SF regime.
    }
    \label{fig:model}
\end{figure}

\vspace{6pt}
\noindent
\textbf{Physical regimes.—}
Figure~\ref{fig:model} illustrates the contrast of a single impurity between a
compressible superfluid (SF) bath and an incompressible
Mott-insulating (MI) bath.
In the SF regime, bath bosons (blue wavy lines)
form a coherent condensate with long-range phase coherence.
A repulsive impurity (red disk) pushes away nearby bath particles,
creating a broad depletion cloud,
while an attractive impurity draws in additional bath density,
forming a smooth accumulation halo.
Because the bath is compressible, these density modulations
extend over many sites and the impurity exhibits coherent motion
with finite winding.
In the MI regime, density fluctuations are frozen and only local distortions survive.
Within the MI regime, for moderate $U_{\mathrm{b}}/t$ the impurity acts as a nearly-free defect for both repulsive and attractive impurity-bath couplings $U_{\mathrm{ib}}/t$, while deeper in the MI (larger $U_{\mathrm{b}}/t$) it turns into a self-trapped defect: a vacancy defect on the repulsive side and a particle defect on the attractive side, both characterized by vanishing impurity winding.

We choose the chemical potential $\mu_{\mathrm{b}}$ such that the homogeneous single-component Bose--Hubbard model (with $U_{\mathrm{ib}}/t=0$) is at unit filling. In the presence of a single impurity the bath is treated in the grand-canonical ensemble, so the total bath occupation deviates from unity only by an amount of order $1/L^{2}$, which is negligible for the thermodynamic properties discussed below. Throughout the paper we therefore refer to this situation as a unit-filled bath.

% =========================================================
\section{Methods and observables}
\label{method}

We employ sign-problem-free multi-species worm-algorithm quantum Monte Carlo (QMC)
in the path-integral representation.  The method extends the original
worm framework of Refs.~\cite{ProkofevJETP1998,secondworm} to two bosonic
species~\cite{PhysRevA.81.053622,Lingua2018}—impurity and bath—and samples their worldline
configurations in imaginary time.  Throughout this work we fix the impurity
number to exactly one, i.e.\ the impurity is simulated in the canonical
ensemble, while the bath bosons are treated in the grand-canonical ensemble
with a chemical potential chosen to stabilize unit filling in the SF and MI
regimes.

The algorithm introduces two independent worm operators that allow
creation, moving, and annihilation of worldline discontinuities on each
species, enabling efficient sampling of both diagonal and off-diagonal
correlators.  

We simulate lattices up to $L=20$ with periodic boundary conditions and use
inverse temperatures $\beta=L$, which is sufficient to reach ground-state
convergence in both the superfluid and the Mott-insulating regimes.

%\subsection{Observables: Impurity-centered measurements}
%\label{observables}

In the following, we summarize the observables used to characterize the impurity mobility and the accompanying bath deformation. All measurements are performed in the (canonical) single-impurity sector and in the grand-canonical ensemble for the bath, using the multi-species worm algorithm. Three global observables are central to our analysis. First, the impurity winding number $\langle W_{\mathrm{imp}}^{2}\rangle$ quantifies the coherent mobility of the impurity: a finite value signals a mobile light polaron, while its continuous suppression and eventual collapse indicate interaction-driven self-trapping, ranging from heavy polarons and saturated bubbles or bound clusters in a superfluid bath to fully self-trapped defects in a Mott-insulating bath. Second, the bath superfluid density $\rho_{\mathrm{b}}$, obtained from the standard winding-number estimator, tracks the loss of global phase stiffness and thus diagnoses the SF--MI transition. Third, impurity-centered density responses, encoded in the correlator
$C_{\mathrm{ib}}(R)$ and its cumulative bath-density deformation
$\Delta N(R)$, resolve the real-space structure of the dressing cloud and help
distinguish depletion bubbles, bound clusters, nearly free defects, and
self-trapped (defect-bound) states. Statistical errors are estimated via
jackknife analysis over independent Monte Carlo blocks.

\subsection{Impurity weight and bath baseline}

The bath density operator is $n_{\mathrm{b}}(\mathbf r)$, and the impurity
occupation operator is $n_{\mathrm{imp}}(\mathbf r)$, which satisfies
$\sum_{\mathbf r}\langle n_{\mathrm{imp}}(\mathbf r)\rangle=1$.
Here $\mathbf r$ labels lattice sites and is equivalent to the site index $i$
used in Sec.~\ref{model}; sums over $\mathbf r$ are therefore equivalent to
sums over lattice sites.
The impurity worldline may visit any site during imaginary time; its
imaginary-time averaged spatial weight is defined as
\begin{equation}
w(\mathbf r_{\mathrm{imp}})
= \frac{1}{\beta}\!\int_{0}^{\beta}
   \langle n_{\mathrm{imp}}(\mathbf r_{\mathrm{imp}},\tau)\rangle\, d\tau,
\qquad \sum_{\mathbf r_{\mathrm{imp}}}w(\mathbf r_{\mathrm{imp}})=1 .
\end{equation}

The uniform bath density (baseline) is
\begin{equation}
\bar n_{\mathrm{b}}
= L^{-2}\sum_{\mathbf r}\langle n_{\mathrm{b}}(\mathbf r)\rangle.
\end{equation}
Spatial displacements are computed using the minimum–image convention under
periodic boundary conditions.

In the worldline representation the impurity can wind around the periodic
boundaries. We define the impurity winding vector
$\mathbf W_{\mathrm{imp}}=(W_{\mathrm{imp}}^x,W_{\mathrm{imp}}^y)$ as the net
number of times the impurity worldline wraps around the system in the $x$ and
$y$ directions, respectively. As a scalar diagnostic we use
\begin{equation}
W_{\mathrm{imp}}^2 \equiv \bigl(W_{\mathrm{imp}}^x\bigr)^2
+ \bigl(W_{\mathrm{imp}}^y\bigr)^2 ,
\end{equation}
and monitor $\langle W_{\mathrm{imp}}^2\rangle$ to distinguish between mobile
and self-trapped impurity regimes.

The bath superfluid density is obtained from the standard winding-number
estimator. Denoting by
$\mathbf W_{\mathrm{b}}=(W_{\mathrm{b}}^x,W_{\mathrm{b}}^y)$ the winding
numbers of the bath worldlines, we define $W_{\mathrm{b}}^2 =
(W_{\mathrm{b}}^x)^2+(W_{\mathrm{b}}^y)^2$ and compute~\cite{PhysRevB.36.8343}
\begin{equation}
\rho_{\mathrm{b}} =
\frac{\langle W_{\mathrm{b}}^2\rangle}{2 t_{\mathrm{b}}\beta}\,,
\label{eq:rhos}
\end{equation}
for a two-dimensional system with lattice spacing set to unity.

The bath compressibility is measured from number fluctuations in the
grand-canonical ensemble,
\begin{equation}
\kappa_{\mathrm{b}}
= \frac{\beta}{L^{2}}
\left( \langle N_{\mathrm{b}}^{2}\rangle
     - \langle N_{\mathrm{b}}\rangle^{2} \right),
\qquad
N_{\mathrm{b}}=\sum_{\mathbf r} n_{\mathrm{b}}(\mathbf r).
\label{eq:kappa}
\end{equation}
Here $\kappa_{\mathrm{b}}$ vanishes in the incompressible MI regime and is
finite in the SF.

Finally, we extract impurity quasiparticle properties from the single-impurity
Green's function measured in real space. Within the worm algorithm we sample
\begin{equation}
G_{\mathrm{imp}}(\mathbf r,\tau)
= \bigl\langle
   a(\mathbf r,\tau)\,
   a^{\dagger}(\mathbf 0,0)
  \bigr\rangle,
\label{eq:G_real}
\end{equation}
where $\mathbf r$ is the spatial separation between the annihilation and
creation operators. The momentum-resolved Green's function is then obtained by
a discrete Fourier transform over the lattice,
\begin{equation}
G_{\mathrm{imp}}(\mathbf k,\tau)
= \sum_{\mathbf r}
  e^{-i\mathbf k\cdot\mathbf r}\,
  G_{\mathrm{imp}}(\mathbf r,\tau).
\label{eq:G_k}
\end{equation}
For sufficiently large imaginary time $\tau$ and low temperature,  
$G_{\mathrm{imp}}(\mathbf{k},\tau)$ is dominated by the lowest-lying impurity (polaron) state at momentum $\mathbf{k}$,
\begin{equation}
G_{\mathrm{imp}}(\mathbf{k},\tau)
\simeq Z_{\mathbf{k}}\, e^{-E_{\mathrm{p}}(\mathbf{k})\,\tau},
\label{eq:G_asympt}
\end{equation}
where $E_{\mathrm{p}}(\mathbf{k})$ is the polaron energy and $Z_{\mathbf{k}}$ is the quasiparticle residue.  
At $\mathbf{k}=\mathbf{0}$, fitting $G_{\mathrm{imp}}(\mathbf{0},\tau)$ to Eq.~\eqref{eq:G_asympt} yields the polaron ground-state energy $E_{\mathrm{p}}(\mathbf{0})$ and residue $Z_{\mathbf{0}}$.

The effective mass is extracted from the small-momentum dispersion,
$m^*/m_0= \frac{2t}{\partial^2 E_{\mathrm{p}}(\mathbf{k})/\partial \mathbf{k}^2},$
where the curvature $\partial^2 E_{\mathrm{p}}(\mathbf{k})/\partial \mathbf{k}^2$ is obtained by fitting the lowest available momenta.

\subsection{Impurity-centered correlator and radial averaging}

We define the impurity-centered correlator as
\begin{align}
G_{\mathrm{ic}}(\mathbf r)
  &= \sum_{\mathbf r_{\mathrm{imp}}} w(\mathbf r_{\mathrm{imp}})
     \langle n_{\mathrm{b}}(\mathbf r_{\mathrm{imp}}+\mathbf r)\rangle ,\\
C_{\mathrm{ib}}(\mathbf r)
  &= G_{\mathrm{ic}}(\mathbf r)-\bar n_{\mathrm{b}}.
\end{align}
Here $\mathbf r_{\mathrm{imp}}$ labels the lattice site occupied by the impurity, $C_{\mathrm{ib}}(\mathbf r)$ measures the change of bath density
\emph{conditioned} on the impurity location:
$C_{\mathrm{ib}}(\mathbf r)<0$ corresponds to depletion (repulsive
$U_{\mathrm{ib}}/t>0$), while $C_{\mathrm{ib}}(\mathbf r)>0$ corresponds to
accumulation (attractive $U_{\mathrm{ib}}/t<0$).

To suppress lattice anisotropy we perform rotational averaging over shells
of fixed Euclidean distance $R=|\mathbf r|$:
\[
C_{\mathrm{ib}}(R)
= \frac{1}{N_R}\sum_{|\mathbf r|=R} C_{\mathrm{ib}}(\mathbf r),
\]
where $N_R$ counts the sites satisfying $|\mathbf r|=R$
(using the minimum–image convention).
This is identical to the definition used in our companion Letter~\cite{chaoletter}.

\subsection{Cumulative excess and its characteristic measures}

The impurity-centered cumulative bath-density deformation,
i.e., the net change of bath particle number within a distance $R$ from
the impurity, is defined as: 
\begin{equation}
\Delta N(R)
  = \sum_{|\mathbf r|\le R} C_{\mathrm{ib}}(\mathbf r)
  = \sum_{R'\le R} N_{R'}\,\overline{C}_{\mathrm{ib}}(R') .
\end{equation}
By construction, $\Delta N(R\!\to\!\infty)=0$ (up to statistical noise), because $C_{\mathrm{ib}}(\mathbf r)$ is defined relative to the global baseline $\bar n_{\mathrm{b}}$ within the same ensemble, while the behavior at small $R$ reveals the local deformation created by the impurity.

In this work, we use two directly measurable quantities extracted from
the cumulative profile.

\paragraph*{(i) Peak radius.}
The radius at which the cumulative excess attains its extremum,
\begin{equation}
R(\Delta N_{\rm ext})
=
\begin{cases}
\arg\min_R\,\Delta N(R), & U_{\mathrm{ib}}/t>0 ,\\[4pt]
\arg\max_R\,\Delta N(R), & U_{\mathrm{ib}}/t<0 ,
\end{cases}
\end{equation}
provides a geometric estimate of the spatial extent of the depletion
bubble for repulsive couplings and of the bound cluster for attractive
couplings.

\paragraph*{(ii) Peak amplitude.}
The extremal value itself,
\begin{equation}
\Delta N_{\rm ext}
=
\begin{cases}
\min_R\,\Delta N(R), & U_{\mathrm{ib}}/t>0 ,\\[4pt]
\max_R\,\Delta N(R), & U_{\mathrm{ib}}/t<0 ,
\end{cases}
\end{equation}
quantifies the number of bath particles expelled or recruited by the
impurity.
For a saturated vacancy bubble, $\Delta N_{\rm ext}\!\approx\!-1$;
for a multi-particle bound cluster, $\Delta N_{\rm ext}\!>\!1$.

These two quantities, $R(\Delta N_{\rm ext})$ and $\Delta N_{\rm ext}$,
are exactly the ones plotted in
Figs.~\ref{fig:att_SF_realspace},~\ref{fig:MIrep1},~\ref{fig:MIatt2},
and~\ref{fig:att_vertical1}.
Together they provide a direct and unambiguous characterization of
depletion-bubble and bound-cluster formation in both SF and MI baths.

\subsection{On-site contrast}

The most local indicator of the deformation is
\begin{equation}
C_{\mathrm{ib}}(\mathbf 0)
 = \langle n_{\mathrm{b}}(\mathbf r_0)\rangle_{\rm imp}
   - \bar n_{\mathrm{b}} ,
\end{equation}
which captures the degree of on-site depletion (repulsive) or
accumulation (attractive).

Together, $\langle W_{\mathrm{imp}}^2\rangle$, $C_{\mathrm{ib}}(R)$,
$\Delta N(R)$, and the extremal measures
$\{R(\Delta N_{\rm ext}),\,\Delta N_{\rm ext}\}$
provide a detailed microscopic characterization of impurity dressing and
self-trapping in both the SF and MI phases and across the SF–MI transition.

%=============================================
% SF 
\section{Attractive impurity in a superfluid bath: real-space self-trapping}
\label{sec:SF}

In our companion Letter~\cite{chaoletter}, we showed that, in a compressible superfluid bath at fixed $U_{\mathrm{b}}/t=13.3$, impurity self-trapping is governed by an interaction-driven winding-collapse mechanism: as a repulsive impurity--bath coupling $U_{\mathrm{ib}}/t>0$ is increased, the impurity evolves from a mobile light polaron with finite winding and an extended dressing cloud to a heavy self-trapped state and, eventually, to a depletion saturated-bubble configuration with $\Delta N(R)\to -1$. In this section we turn to the attractive side. To reveal the mirror counterpart of the repulsive depleted bubble and to resolve the real-space structure of self-trapping, we again fix the bath interaction at $U_{\mathrm{b}}/t=13.3$ (in the superfluid regime) and tune the attractive impurity--bath coupling $U_{\mathrm{ib}}/t<0$.

\begin{figure*}[t]
  \centering
  \includegraphics[width=\textwidth]{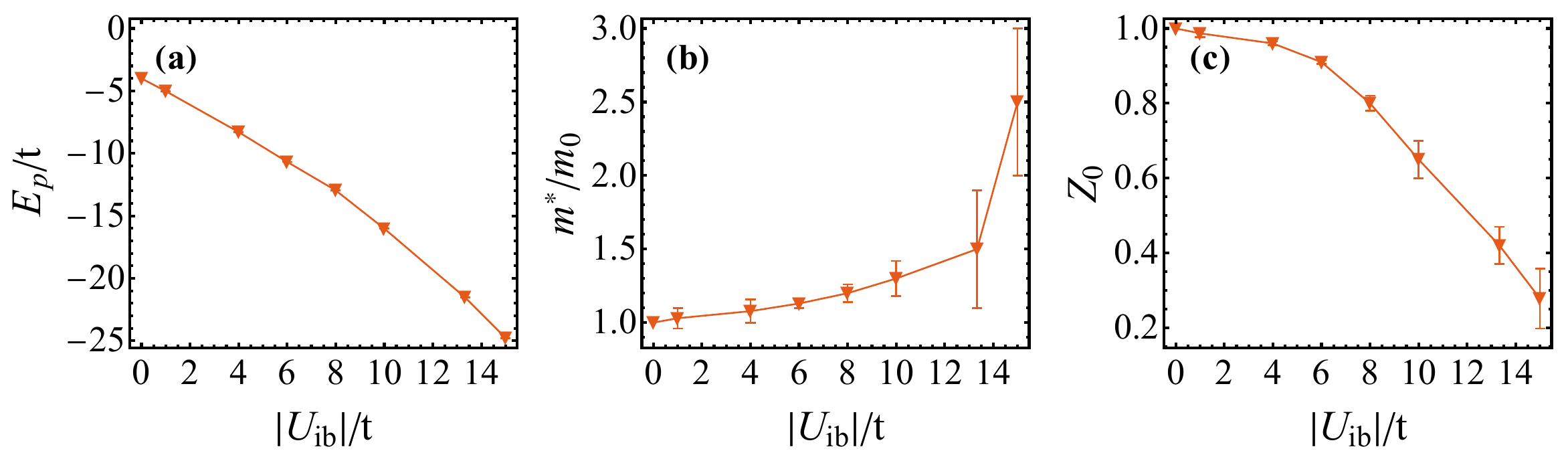}
  \caption{\textbf{Quasiparticle properties of an attractive impurity in a superfluid bath.}
  (a) the polaron ground-state energy $E_{\mathrm{p}}/t$, (b) the effective-mass
  ratio $m^{*}/m_0$ with $m_0$ the bare lattice mass of the impurity, and (c) the quasiparticle residue $Z_0$ as functions of the
  attractive impurity--bath coupling $|U_{\mathrm{ib}}|/t$ at fixed bath
  interaction $U_{\mathrm{b}}/t=13.3$, deep in the superfluid regime.
  Error bars denote statistical uncertainties and are smaller than the symbols
  where not visible.
  }
  \label{fig:SF_att_quasi}
\end{figure*}

\begin{figure}[t]
    \centering
\includegraphics[width=\linewidth, trim = 0pt 0pt 1100pt 0pt,  clip]{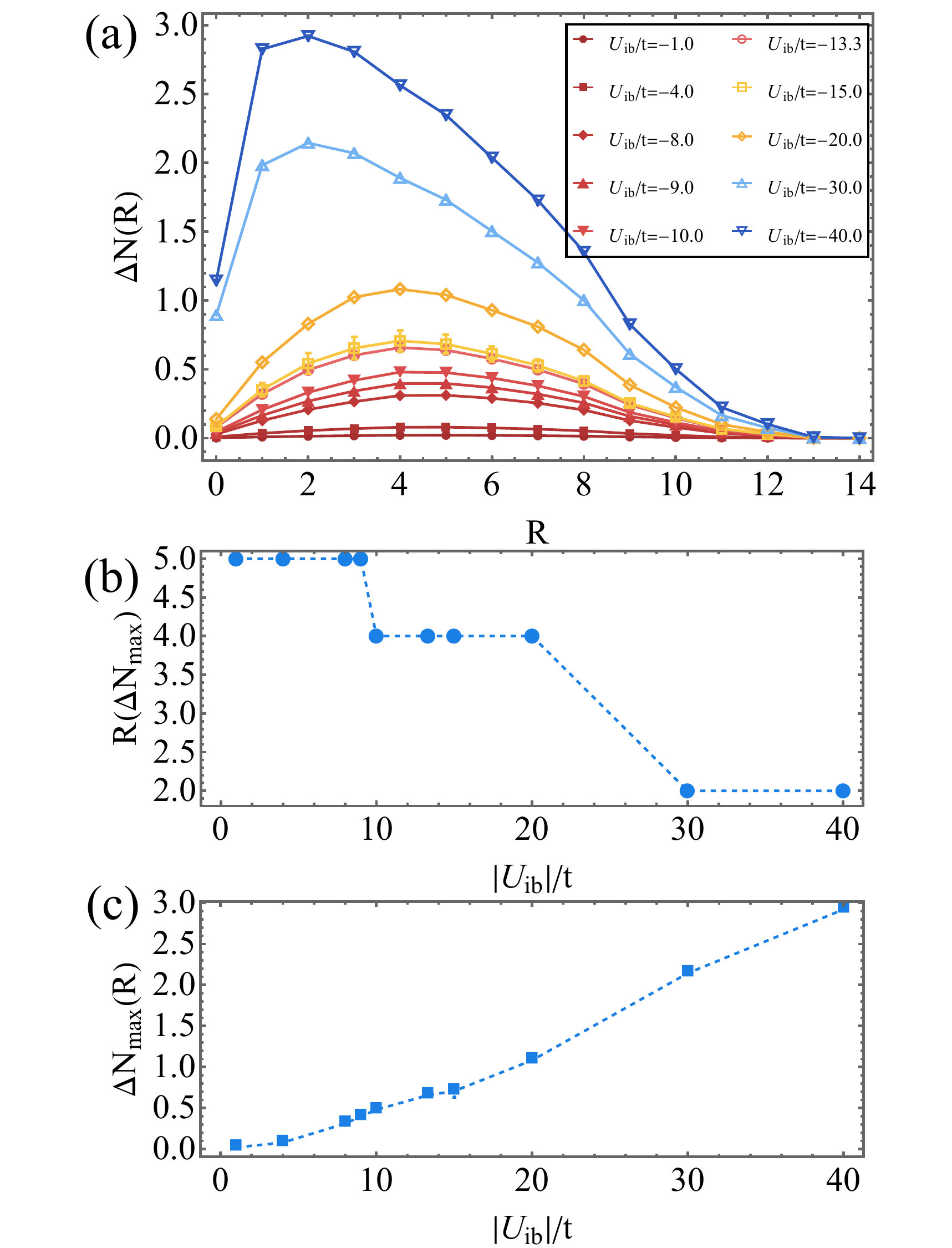}
    \caption{\textbf{Real-space impurity dressing for attractive impurity in a superfluid bath at $U_{\mathrm{b}}/t = 13.3$.} (a) Cumulative bath-density enhancement $\Delta N(R)$ for various attractive impurity--bath couplings $U_{\mathrm{ib}}/t<0$, showing the evolution from a broad, weak dressing cloud (light polaron) to a sharply self-trapped multi-particle cluster.
    (b) Radius of the maximum $R(\Delta N_{\max})$, demonstrating the monotonic inward compression of the dressing cloud as $|U_{\mathrm{ib}}|/t$ increases.
    (c) Magnitude of maximum $\Delta N_{\max}(R)$, which increases without saturation, reflecting the grand-canonical nature of the bath: stronger attraction continuously recruits additional bosons into the impurity vicinity.
    Unlike the repulsive saturated-bubble case where $\Delta N(R)$ saturates at $-1$, the attractive branch exhibits unbounded growth due to unrestricted local accumulation.
    These results demonstrate a continuous crossover:
    light polaron $\rightarrow$ heavy polaron $\rightarrow$ bound cluster,
    while the bath remains globally superfluid.}
    \label{fig:att_SF_realspace}
\end{figure}

\begin{figure}[t]
    \centering
\includegraphics[width=\linewidth, trim = 0pt 50pt 430pt 90pt,  clip]{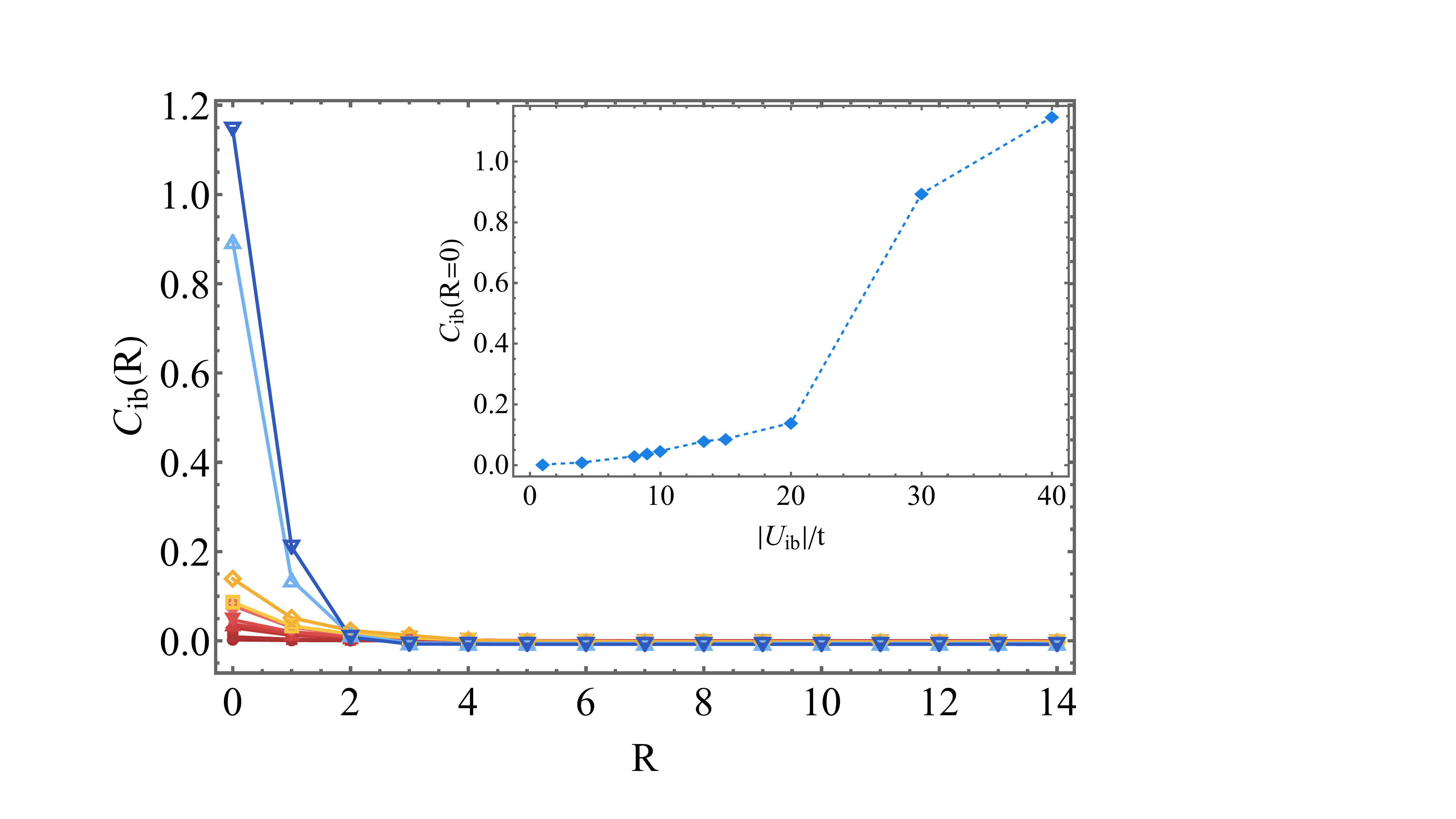}
    \caption{\textbf{Real-space impurity-centered correlator 
    $C_{\mathrm{ib}}(R)$ in the superfluid bath 
    at $U_{\mathrm{b}}/t=13.3$ for various attractive couplings 
    $U_{\mathrm{ib}}/t<0$.} For weak attraction ($|U_{\mathrm{ib}}|/t\lesssim13.3$, red curves),
    $C_{\mathrm{ib}}(R)$ is small and spatially broad,
    indicating an extended, compressible dressing cloud
    characteristic of a mobile light polaron.
    In the intermediate regime 
    ($13.3< |U_{\mathrm{ib}}|/t \lesssim 20.0$, orange curves),
    the response becomes stronger and short-ranged,
    signaling a compact heavy polaron with reduced coherent motion.
    For strong attraction ($|U_{\mathrm{ib}}|/t > 20.0$, blue curves),
    the correlation collapses to a sharply peaked on-site form,
    forming a self-trapped, multi-particle bound cluster.
    The inset shows that the on-site value $C_{\mathrm{ib}}(0)$
    grows monotonically with $|U_{\mathrm{ib}}|/t$ and does not saturate,
    highlighting that in a grand-canonical superfluid bath,
    a deeper attractive impurity continuously recruits additional bath
    bosons without intrinsic upper bound.
    This demonstrates an interaction-driven winding-collapse crossover in a compressible superfluid host.}
    \label{fig:attSF_FIG2}
\end{figure}

\begin{figure}[t]
\centering
\includegraphics[width=\linewidth, trim = 300pt 10pt 300pt 40pt, clip]{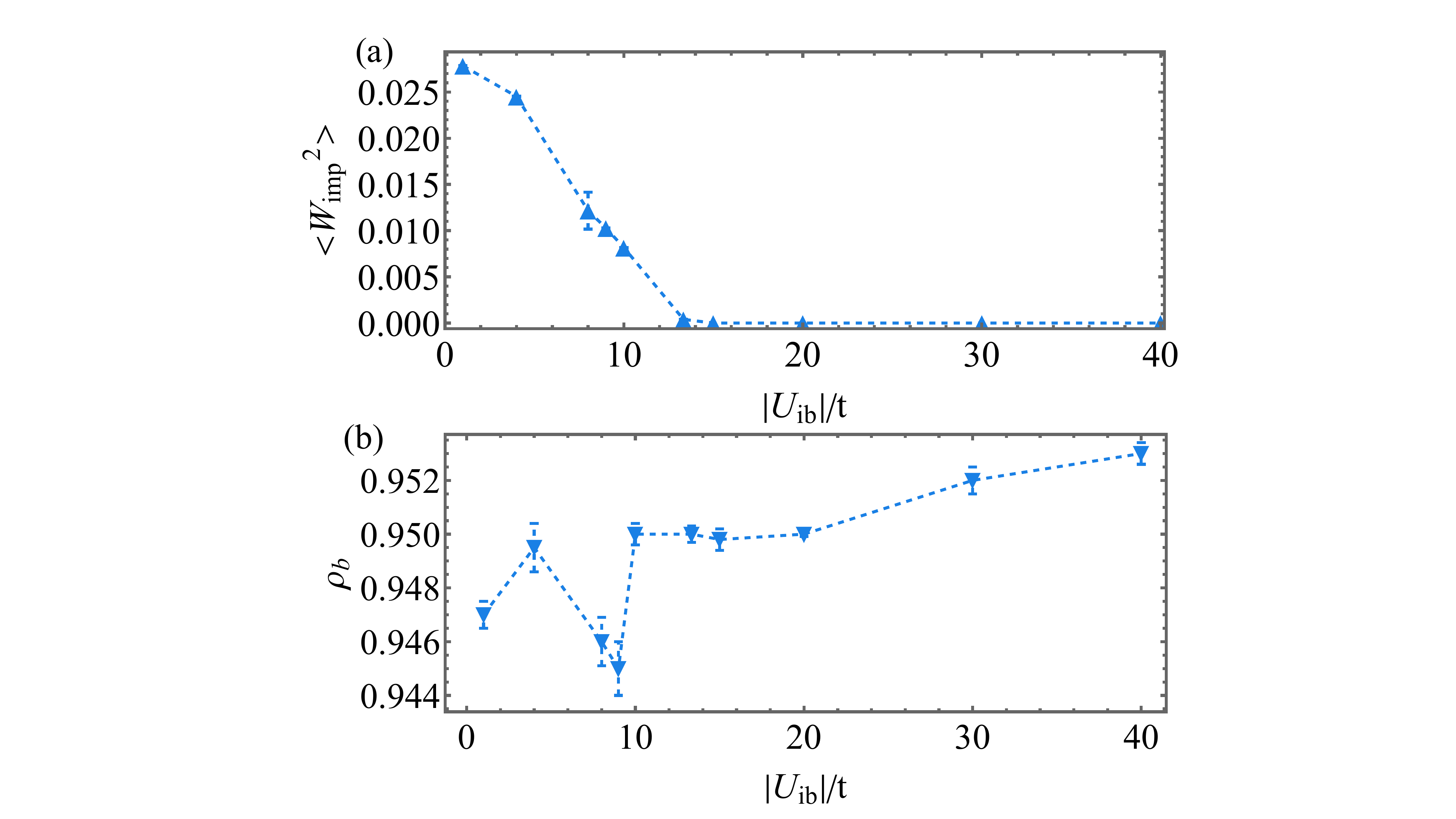}
\caption{\textbf{Interaction-controlled winding-collapse self-trapping of an attractive impurity in a superfluid bath at fixed $U_{\mathrm{b}}/t=13.3$.} 
(a) Impurity winding number $\langle W_{\mathrm{imp}}^{2}\rangle$ as a function of 
attractive coupling $U_{\mathrm{ib}}/t<0$. 
A smooth suppression followed by a rapid collapse at $|U_{\mathrm{ib}}|/t\gtrsim10$ 
signals a transition from a mobile light polaron to a self-trapped bound cluster.
(b) Bath superfluid density $\rho_{\mathrm{b}}$ remains essentially unchanged across 
the entire coupling range, confirming that the impurity self-trapping is \emph{not} 
induced by the loss of global coherence in the bath.
This establishes an interaction-driven winding-collapse self-trapping mechanism: 
the impurity becomes self-trapped even while the bath remains fully superfluid 
and compressible.}
\label{fig:attSF_winding}
\end{figure}

\paragraph*{Quasiparticle properties along an interaction-driven trajectory.}
We start with the interaction-driven path in the superfluid bath and track
how the impurity quasiparticle evolves as the attractive impurity-bath coupling is increased
at fixed $U_{\mathrm{b}}/t=13.3$. Figure~\ref{fig:SF_att_quasi} shows the
polaron energy $E_{\mathbf{p}}/t$, the effective-mass ratio $m^{*}/m_0$, and the residue $Z_0$
as functions of $U_{\mathrm{ib}}/t<0$. For weak attraction $|U_{\mathrm{ib}}|/t<10.0$ the impurity forms a
moderately dressed light polaron: the energy is only linearly shifted, the mass
remains close to the bare value ($m^{*}/m_0\approx1$–$1.2$), and the residue is
high, $Z_0\approx1.0$–$0.6$. As $|U_{\mathrm{ib}}|/t$ is increased, the energy
drops almost linearly, the effective mass grows steadily to $m^{*}/m_0\gtrsim1.5$,
and the residue is reduced to $Z_0\sim0.2$, indicating progressively stronger
dressing. Upon entering the strong-attraction regime $|U_{\mathrm{ib}}|/t>13.3$, the mass shows a sharp upturn while the residue is strongly suppressed, signaling the heavy polaron formation with very limited coherent mobility.
This quasiparticle evolution is fully consistent with the winding-number
collapse and the real-space contraction of the dressing cloud discussed in
Figure~\ref{fig:att_SF_realspace} to~\ref{fig:attSF_winding}, and together they establish a continuous crossover from a
mobile light polaron at weak attractive impurity-bath coupling $|U_{\mathrm{ib}}|/t$ to a heavy polaron, finally a nearly self-trapped bound cluster at strong attraction in a still globally superfluid bath.

\paragraph*{Attractive-branch dressing in a compressible superfluid bath.}
Figure~\ref{fig:att_SF_realspace} summarizes the real-space characteristics of the impurity dressing in the attractive branch, obtained at fixed bath interaction $U_{\mathrm{b}}/t=13.3$, well inside the superfluid regime. At weak coupling ($|U_{\mathrm{ib}}|/t\lesssim 13.3$), 
$\Delta N(R)$ exhibits a broad, small-amplitude hump, indicating a light Bose polaron characterized by an extended dressing cloud and finite impurity winding. As the attraction strengthens ($13.3\lesssim|U_{\mathrm{ib}}|/t\lesssim20.0$), the peak amplitude $\Delta N_{\max}(R)$ grows, signaling a compression of the dressing cloud and the emergence of a heavy-polaron regime with reduced coherent mobility and nearly zero windings.

For stronger attractive impurity--bath coupling ($|U_{\mathrm{ib}}|/t>20.0$),
the deformation cloud collapses toward the impurity site and
$\Delta N(R)$ develops a sharply peaked profile at $R\simeq 2$,
indicating the formation of a tightly bound multi-particle cluster.
Unlike the repulsive saturated-bubble case, where the cumulative deficit
$\Delta N(R)$ rapidly saturates at approximately $-1$ once $R$ exceeds
the depletion radius, here the cumulative excess
$\Delta N_{\max}$ \emph{continues to grow} with $|U_{\mathrm{ib}}|/t$.
Both branches are simulated in the same grand-canonical ensemble for
the bath; the qualitative difference originates from the energetics of
adding versus removing particles.
For a repulsive impurity it is energetically favorable to expel essentially
one bath boson from the impurity vicinity, producing a saturated depletion
bubble with $\Delta N_{\min}\simeq -1$.
For attractive $U_{\mathrm{ib}}/t<0$, the local potential well can instead
bind several bath bosons: as $|U_{\mathrm{ib}}|/t$ increases the optimal
on-site occupation near the impurity grows to $C_{\rm ib}(R=0)\approx 3\text{--}4$
for our strongest couplings, so the bound cluster recruits a few extra
particles from the reservoir.
In a $20\times 20$ system this corresponds to a total bath occupation
of $N_{\mathrm{b}}\simeq 403$, i.e., a global filling only slightly above
unity, while the local excess is fully captured by the cumulative bath-density response
$\Delta N(R)$.
Taken together, these results establish the continuous crossover within
the superfluid bath, $\textit{light polaron} \;\rightarrow\; \textit{heavy polaron}
\;\rightarrow\; \textit{bound cluster}$, all occurring while the bath remains globally superfluid.

In Fig.~\ref{fig:attSF_FIG2}, we examine the impurity-centered correlation function $C_{\mathrm{ib}}(R)$, which provides real-space insight into how the impurity distorts its local environment. The bath interaction is fixed at $U_{\mathrm{b}}/t=13.3$, well inside the superfluid phase, while the attractive impurity--bath coupling $U_{\mathrm{ib}}/t<0$ is tuned from weak to strong. For weak attraction ($|U_{\mathrm{ib}}|/t\lesssim13.3$, red curves), $C_{\mathrm{ib}}(R)$ displays a shallow and spatially broad profile, reflecting a compressibility-enabled, extended dressing cloud characteristic of a coherent light polaron with finite winding and significant mobility, consistent with Fig.~\ref{fig:attSF_winding}(a). In the intermediate regime ($13.3 < |U_{\mathrm{ib}}|/t\lesssim20$, orange curves), the correlation grows in amplitude while the winding has already collapsed to zero, signaling the formation of a compact heavy-polaron state with strongly reduced coherent motion. For strong attraction ($|U_{\mathrm{ib}}|/t \gtrsim 20$, blue curves), all curves have a dominant on-site peak and a pronounced nearest-neighbor maximum, indicative of a tightly bound multi-particle cluster. The inset shows that the on-site signal $C_{\mathrm{ib}}(0)$ increases \emph{monotonically} with $|U_{\mathrm{ib}}|/t$ and does not saturate. This behavior reflects the grand-canonical nature of the superfluid bath at fixed chemical potential: a deeper attractive potential can continuously recruit additional bath bosons into the cluster. It contrasts sharply with the repulsive branch, where the cumulative bath-density response $\Delta N(R)$ saturates at approximately $-1$ once a single bath particle has been expelled from the impurity vicinity.

Taken together, these results demonstrate a continuous, interaction-driven winding-collapse crossover from a coherent light polaron to a heavy polaron and finally a self-trapped, multi-particle bound cluster, even though the bath remains globally superfluid. The collapse of the spatial dressing profile thus provides a real-space signature of interaction-driven self-trapping at fixed bath compressibility.

\paragraph*{Winding-controlled self-trapping of an attractive impurity in the superfluid bath.} To directly probe how impurity mobility degrades with increasing attractive coupling, 
Fig.~\ref{fig:attSF_winding} summarizes the evolution of the impurity winding number 
$\langle W_{\mathrm{imp}}^{2}\rangle$ and the bath superfluid density $\rho_{\mathrm{b}}$ 
as a function of $U_{\mathrm{ib}}/t<0$ at fixed $U_{\mathrm{b}}/t=13.3$, deep in the 
compressible superfluid regime.

Figure~\ref{fig:attSF_winding}(a) shows that $\langle W_{\mathrm{imp}}^{2}\rangle$ decreases smoothly at first, 
then rapidly collapses by nearly two orders of magnitude once 
$|U_{\mathrm{ib}}|/t\gtrsim 13.3$, signaling a crossover from a mobile light polaron to 
a heavy polaron. 
This loss of winding takes place without any qualitative change in the bath, as confirmed 
in Fig.~\ref{fig:attSF_winding}~(b), where the bath superfluid density $\rho_{\mathrm{b}}$ remains essentially 
constant across the entire range of impurity-bath couplings.
The coexistence of vanishing impurity winding with a finite bath superfluid density
clearly demonstrates that self-trapping is \emph{not} driven by the breakdown of
global phase coherence in the bath, but rather by an \emph{interaction-driven
winding collapse}, i.e., self-trapping of the impurity.

In contrast to a Mott-insulating background, where impurity self-trapping involves quantized defect 
formation and abrupt expulsion or recruitment of a single bath boson, here the collapse 
of mobility is continuous. 
The bath remains fully superfluid and compressible, yet the impurity loses all 
long-range coherence and becomes self-trapped purely by increasing the depth of the 
local attractive impurity-bath coupling. 
This establishes a clear case of \emph{interaction-driven winding-collapse self-trapping in a superfluid bath}:
the impurity loses its ability to form worldline windings even though the bath retains 
macroscopic superfluidity.

\section{Impurity in a Mott-insulating bath}
\label{sec:MI}

Having established how the impurity evolves when the bath is in SF, we now turn
to the complementary question: \emph{What is the nature of the impurity once
the bath is already deep in the Mott-insulating phase?}  
In this incompressible regime, long-wavelength density fluctuations are
frozen and the superfluid density vanishes, removing the screening
mechanisms that allow extended polarons to form in a compressible
superfluid.  Consequently, the impurity can perturb its environment only
through strictly local rearrangements of bath occupation, in sharp
contrast to the smooth, extended bath density deformations found in the SF
regime.

A key feature of the Mott-insulator state is that bath occupation can change only
in \emph{integer} units.  This quantization strongly constrains the
impurity response: under a repulsive impurity-bath coupling ($U_{\mathrm{ib}}/t>0$), the
impurity may lower its energy only by \emph{expelling one} bath boson from
its vicinity; under an attractive impurity-bath coupling ($U_{\mathrm{ib}}/t<0$), it may
\emph{recruit one} additional boson.  Extended deformations, multi-particle
halos, and continuous dressing clouds are forbidden by the
incompressibility of the bath.  Thus, impurity self-trapping in the MI
proceeds via discrete $\pm 1$–particle defects rather than via
compressibility-enabled screening.

In the following two subsections we analyze these two cases
separately at fixed bath interaction $U_{\mathrm{b}}/t=20.0$, representative of the deep Mott-insulating regime.
For repulsive impurity-bath coupling, we show that the impurity evolves from a
weakly dressed, nearly-free defect at small $U_{\mathrm{ib}}/t$ to a fully
self-trapped \emph{vacancy defect} in which exactly one bath boson is expelled.
For attractive impurity-bath coupling, we find an analogous progression from a
lightly dressed defect to a self-trapped \emph{particle defect} that binds one
additional boson.  

Together, these two trajectories establish the characteristic,
quantized signatures of impurity self-trapping in an incompressible bath
and highlight their fundamental difference from the continuous,
bath interaction-driven dressing found on the superfluid side.

% ===================== FIG MIrep1 ============================
\begin{figure}[t]
\centering
\includegraphics[width=\linewidth, trim =0pt  0pt 1100pt 0pt, clip]{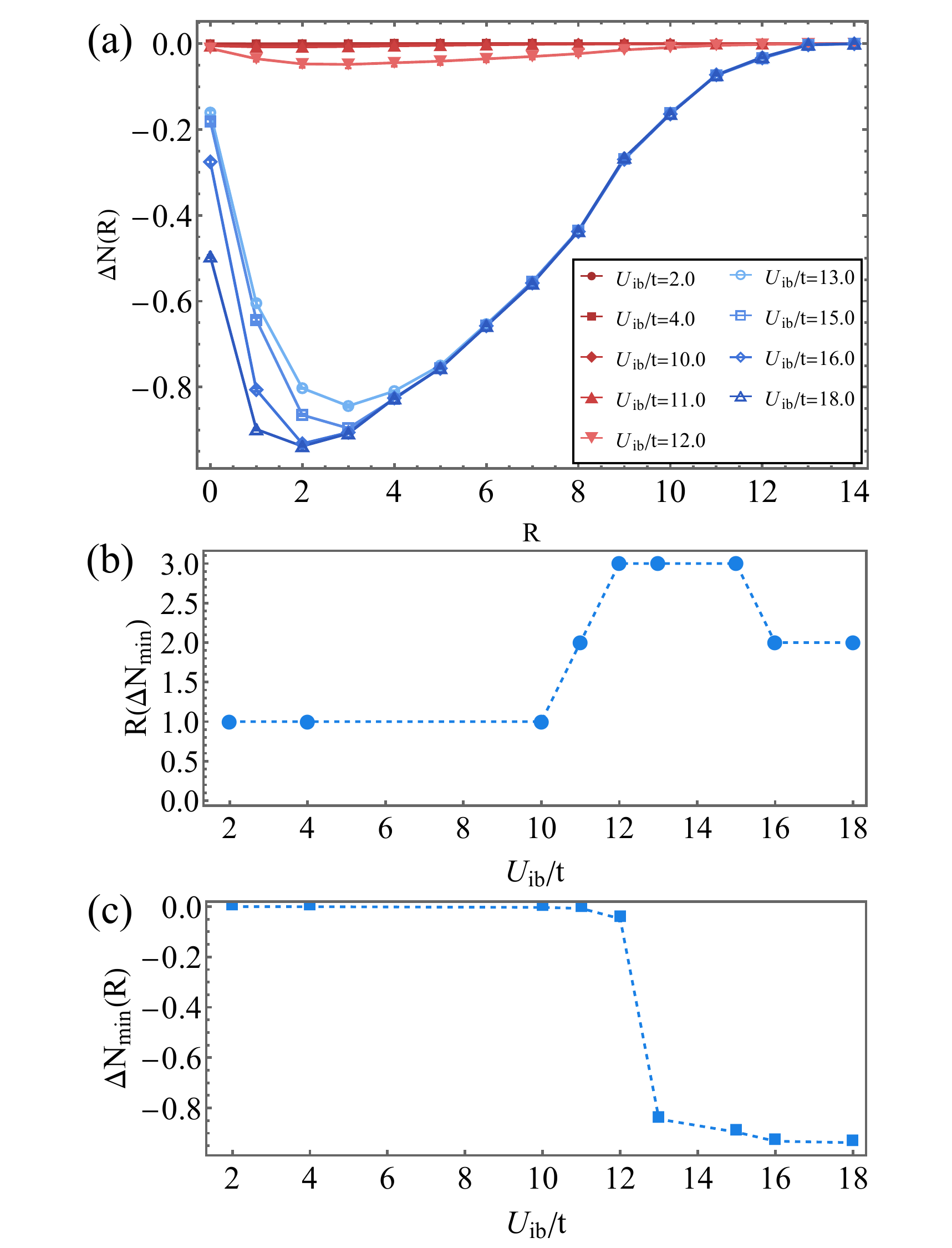}
\caption{\textbf{Repulsive impurity in a Mott-insulating bath: cumulative bath-density deformation.}
(a)~Cumulative bath-density deformation 
$\Delta N(R)$
for increasing repulsive coupling $U_{\mathrm{ib}}/t$ at $U_{\mathrm{b}}/t=20.0$.
For weak $U_{\mathrm{ib}}/t$, the response shows only a shallow minimum,
indicative of a nearly-free defect.
For $U_{\mathrm{ib}}/t\!\gtrsim\!13.0$, the minimum approaches 
$\Delta N_{\min}\approx -1$, signaling the expulsion of a single bath particle.
(b)~Radius of the minimum, $R(\Delta N_{\min})$, which sharply increases at the
self-trapping threshold.
(c)~Magnitude of the minimum $|\Delta N_{\min}(R)|$, saturating to unity once the
vacancy defect is formed.}
\label{fig:MIrep1}
\end{figure}

% ===================== FIG MIrep2 ============================
\begin{figure}[b]
\centering
\includegraphics[width=\linewidth, trim = 200pt 30pt 200pt 60pt, clip]{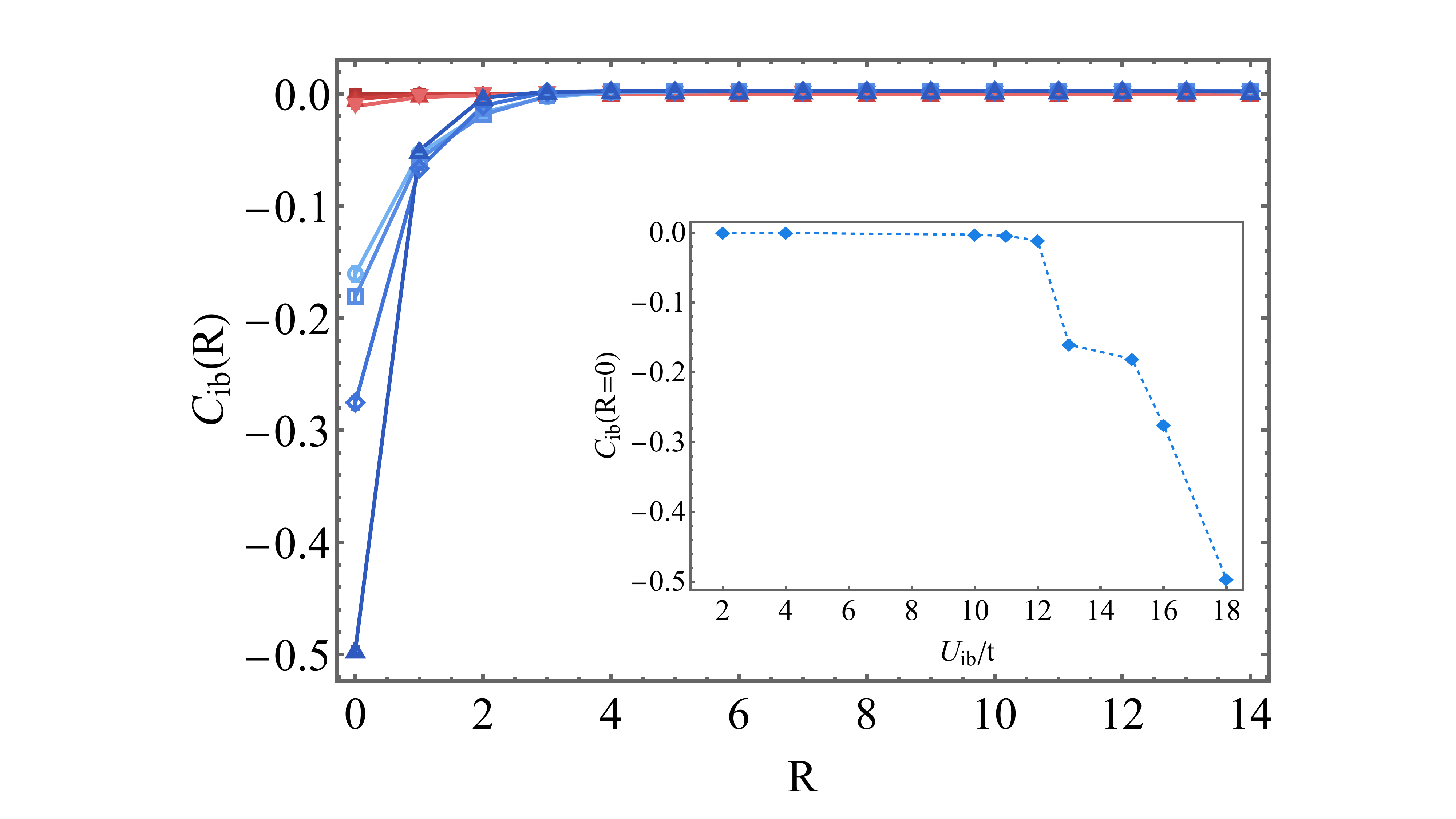}
\caption{\textbf{Repulsive impurity in a Mott-insulating bath: radial impurity--centered correlator.}
Radial-averaged impurity-centered correlator $C_{\mathrm{ib}}(R)$ evolves from a shallow, short-range dip at weak
$U_{\mathrm{ib}}/t$ (nearly-free defect) to a deep on-site depletion for $U_{\mathrm{ib}}/t\!\gtrsim\!12.0$,
indicating formation of a saturated vacancy defect.
\textit{Inset:} on-site contrast $C_{\mathrm{ib}}(0)$, showing a sharp drop at the same
threshold, consistent with the expulsion of one bath particle from the impurity
site.}
\label{fig:MIrep2}
\end{figure}

\subsection{Repulsive impurity in a Mott-insulating bath}
\label{subsec:MI_repulsive}

\begin{figure}[h]
\centering
\includegraphics[width=\linewidth, trim = 200pt 10pt 500pt 40pt, clip]{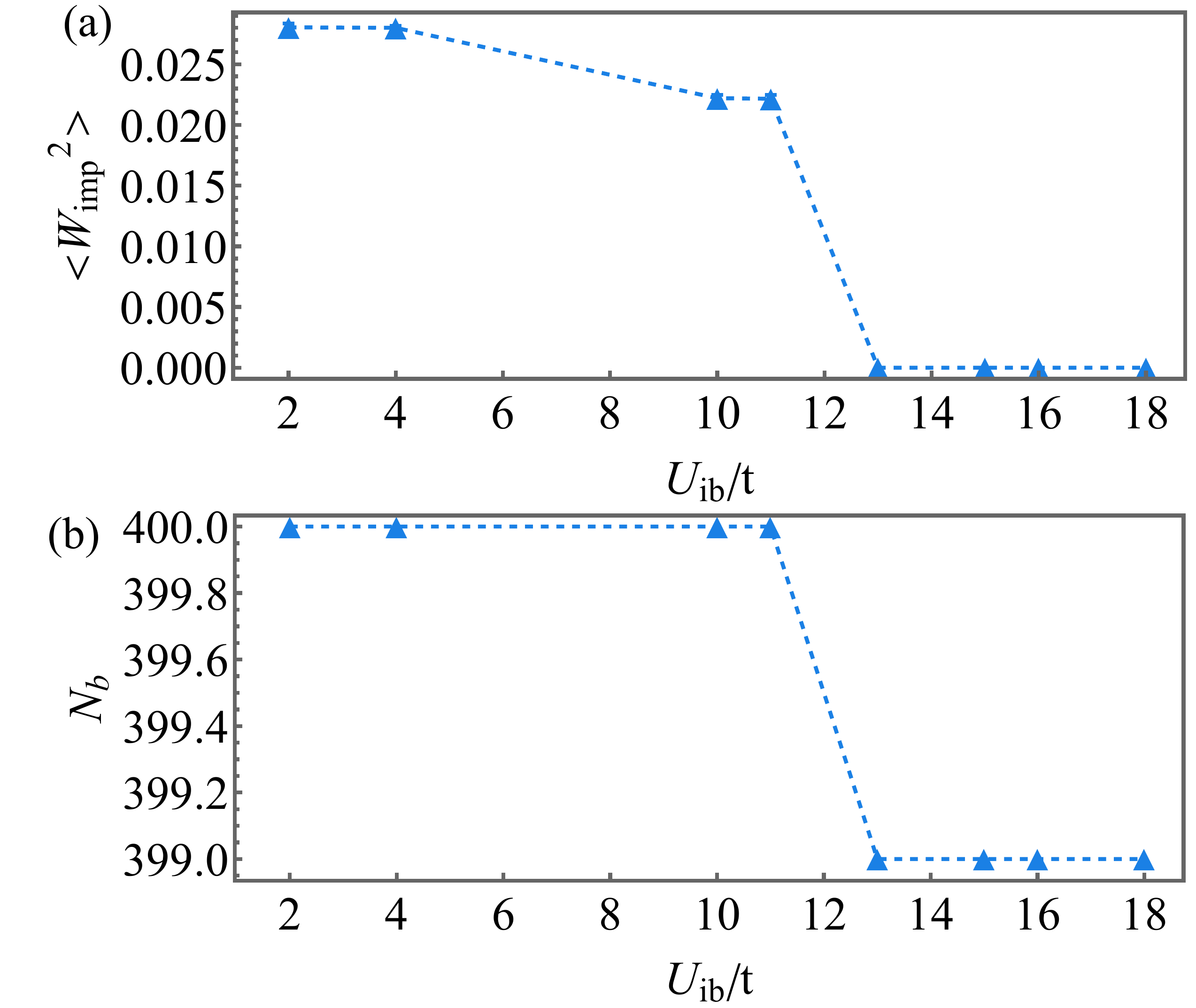}
\caption{\textbf{Impurity coherence and bath particle number for
repulsive impurity-bath coupling in the MI.}
(a)~Impurity winding number $\langle W_{\mathrm{imp}}^2\rangle$.
Finite winding at small $U_{\mathrm{ib}}/t$ confirms coherent motion of a 
nearly-free defect, whereas a rapid collapse when $U_{\mathrm{ib}}/t > 13.0$
marks self-trapping of the impurity--vacancy composite.
(b)~Total bath occupation $N_{\mathrm{b}}$,
showing a quantized decrease by exactly one particle once the vacancy defect
is formed.  
This discrete change highlights that self-trapping in the Mott-insulating regime proceeds by
defect binding rather than extended density redistribution.}
\label{fig:MIrep3}
\end{figure}

We now investigate the behavior of a repulsive impurity
($U_{\mathrm{ib}}/t>0$) embedded in a deep Mott-insulating bath with
$U_{\mathrm{b}}/t=20.0$.  
In this regime, the background is incompressible, long-wavelength density
fluctuations are frozen, and only local particle--hole excitations are accessible.
Under these conditions the impurity cannot form an extended polaronic cloud:
instead, its dressing is strictly local, and self-trapping occurs through 
\emph{discrete defect formation}.  
Figures~\ref{fig:MIrep1}–\ref{fig:MIrep3} summarize our analysis.

\paragraph*{Weak repulsion: nearly-free defect with minimal dressing.}
For repulsive impurity-bath couplings $U_{\mathrm{ib}}/t\!\lesssim\!12.0$, the impurity behaves as a 
mobile, nearly-free defect propagating through the rigid Mott-insulating background.
Figure~\ref{fig:MIrep1}(a) shows the cumulative bath response 
$\Delta N(R)$:
the minimum $\Delta N_{\min}$ is extremely shallow 
($|\Delta N_{\min}|\ll 0.02$), and its radius lies at 
$R(\Delta N_{\min})\approx 1$–$2$ sites,  
as quantified in Fig.~\ref{fig:MIrep1}(b).
The impurity-centered correlator $C_{\mathrm{ib}}(R)$  
[Fig.~\ref{fig:MIrep2}(a)] exhibits only a small on-site dip with no visible
tails, reflecting the short-range, weak rearrangement available in an
incompressible background.
The on-site contrast $C_{\mathrm{ib}}(0)$ [inset of Fig.~\ref{fig:MIrep2}]
remains near zero.
Consistent with this picture, the impurity winding number 
$\langle W_{\mathrm{imp}}^2\rangle$  
remains large and changes less than $10\%$ [Fig.~\ref{fig:MIrep3}(a)],
demonstrating that the impurity retains coherent mobility
even in the absence of superfluidity of the bath.

\paragraph*{Strong repulsion: formation of a self-trapped vacancy defect.}
As $U_{\mathrm{ib}}/t$ increases beyond approximately $13.0$, a sharp restructuring 
occurs in all observables.
The cumulative bath-density deformation develops a deep, quantized minimum  
$\Delta N_{\min}\approx -1$ [Fig.~\ref{fig:MIrep1}(c)],
and the radius $R(\Delta N_{\min})$ jumps to a larger value 
$\approx 2$ or $3$ [Fig.~\ref{fig:MIrep1}(b)].
These two signatures indicate that the impurity \emph{expels exactly one}
bath boson from its vicinity.
The radial impurity-centered correlator $C_{\mathrm{ib}}(R)$ develops a strongly negative
on-site value and a small compensating positive overshoot at larger $R$ 
[Fig.~\ref{fig:MIrep2}(a)]. %, enforcing the sum rule $\sum_{\mathbf r} C_{\mathrm{ib}}(\mathbf r)=0$.
Importantly, neither the magnitude $|\Delta N_{\min}|$ nor the overall 
profile changes with further increase of $U_{\mathrm{ib}}/t$:
the vacancy-defect formation is strictly quantized and fully saturated.

\paragraph*{Coherence collapse and quantized particle loss.}
The vacancy-defect formation is accompanied by a collapse of coherent
motion.  
The winding number $\langle W_{\mathrm{imp}}^2\rangle$  
drops rapidly once $U_{\mathrm{ib}}/t \simeq 13.0$  
and approaches zero for larger impurity-bath couplings
[Fig.~\ref{fig:MIrep3}(a)], signaling complete self-trapping of the
impurity–vacancy composite object.  
Simultaneously, the total bath particle number  
$N_{\mathrm{b}}$  
decreases by \emph{exactly one} particle  
[Fig.~\ref{fig:MIrep3}(b)], 
confirming that the impurity binds to a vacancy in the Mott-insulating background.
This behavior is fundamentally distinct from the superfluid case discussed
in companion:  
there, self-trapping proceeds through a continuous redistribution of density,
while in the Mott-insulating regime it is enforced by the discrete and gapped nature
of particle–hole excitations.  
The impurity therefore transitions from a nearly-free defect to a 
fully self-trapped vacancy defect as $U_{\mathrm{ib}}/t$ increases, providing
a demonstration of defect-mediated self-trapping in an 
incompressible quantum state.

% ===================== FIG MIrep3 ============================

%====================  MI: Attractive impurity subsection  ====================%
\subsection{Attractive impurity in a Mott-insulating bath}
\label{subsec:MI_attractive}

\begin{figure}[t]
    \centering
\includegraphics[width=\linewidth, trim = 10pt 0pt 1100pt 5pt, clip]{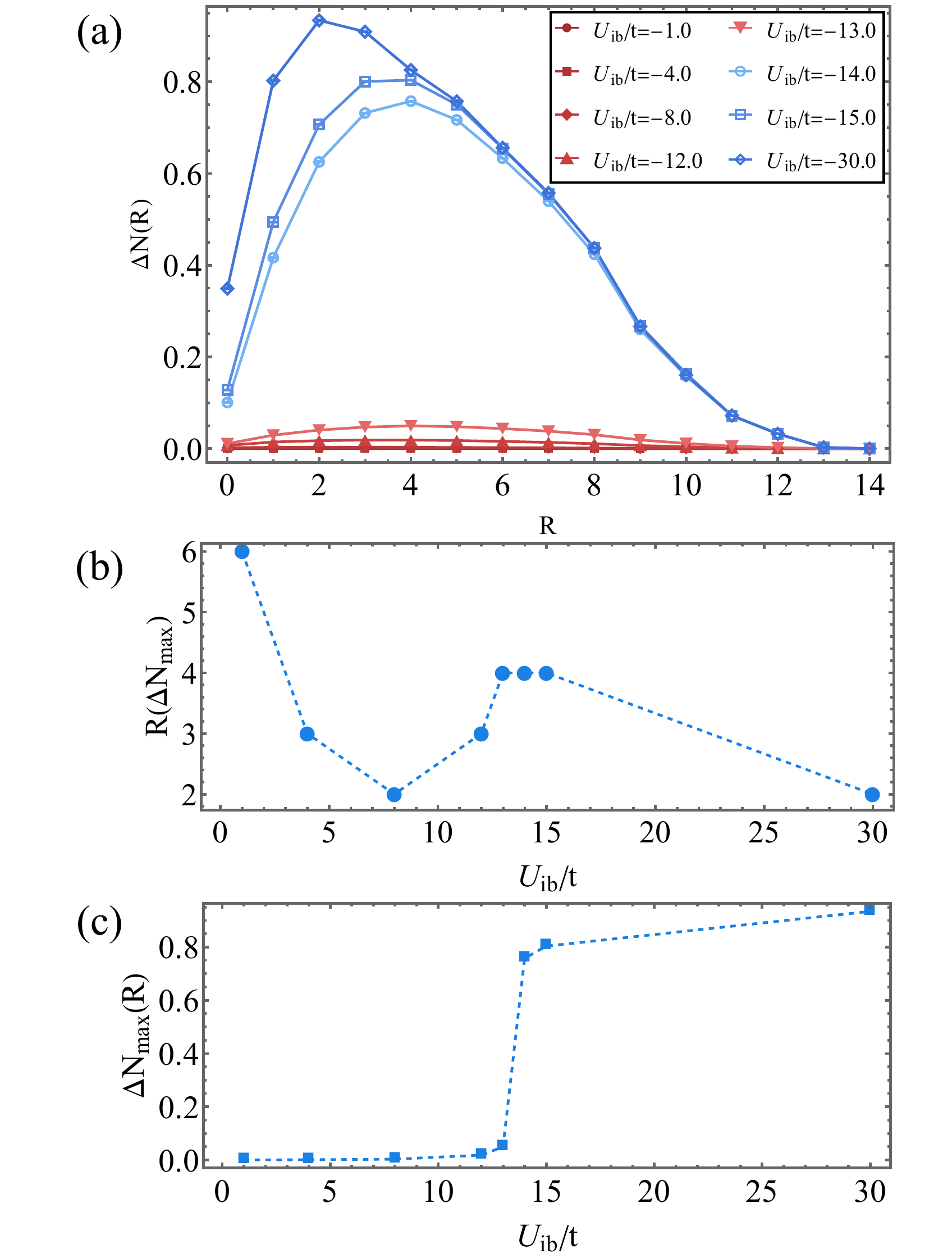}
    \caption{\textbf{Attractive impurity in a Mott-insulating bath: cumulative bath-density deformation.}
    (a)~Cumulative bath-density excess $\Delta N(R)$ for increasing attraction
    $|U_{\mathrm{ib}}|/t$ at $U_{\rm{b}}/t=20.0$, showing the growth of a positive deformation near
    the impurity.
    (b)~Radius of maximum,
    $R(\Delta N_{\max})$,
    extracted from panel~(a).
    (c)~Magnitude of maximum $\Delta N_{\max}(R)$, saturating to unity once the particle defect is formed.}
    \label{fig:MIatt1}
\end{figure}

\begin{figure}[t]
    \centering
\includegraphics[width=\linewidth, trim = 120pt 80pt 300pt 80pt, clip]{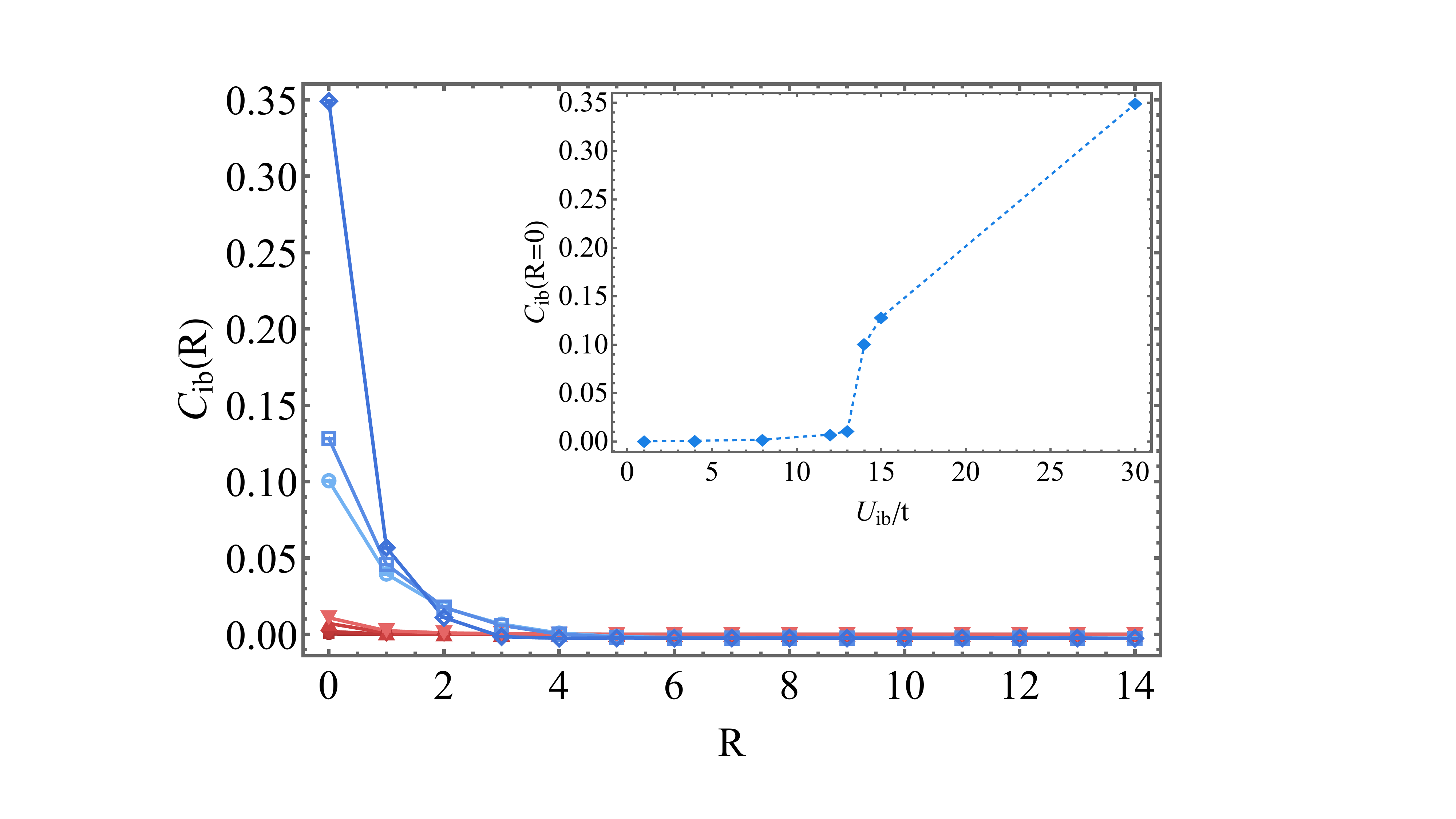}
    \caption{\textbf{Attractive impurity in a Mott-insulating bath: radial impurity-centered correlator.}
    Radially averaged impurity-centered correlator $C_{\mathrm{ib}}(R)$
    for various $|U_{\mathrm{ib}}|/t$,
    showing the evolution from a weak local enhancement
    (nearly-free defect) to a strong, sharply self-trapped peak
    (particle-defect).
    \emph{Inset:}
    on-site contrast $C_{\mathrm{ib}}(0)$ versus $U_{\mathrm{ib}}/t$,
    illustrating the abrupt growth of local density once the
    particle-defect formation threshold is crossed.}
    \label{fig:MIatt2}
\end{figure}

\begin{figure}[b]
    \centering
\includegraphics[width=\linewidth, trim = 200pt 10pt 360pt 40pt, clip]{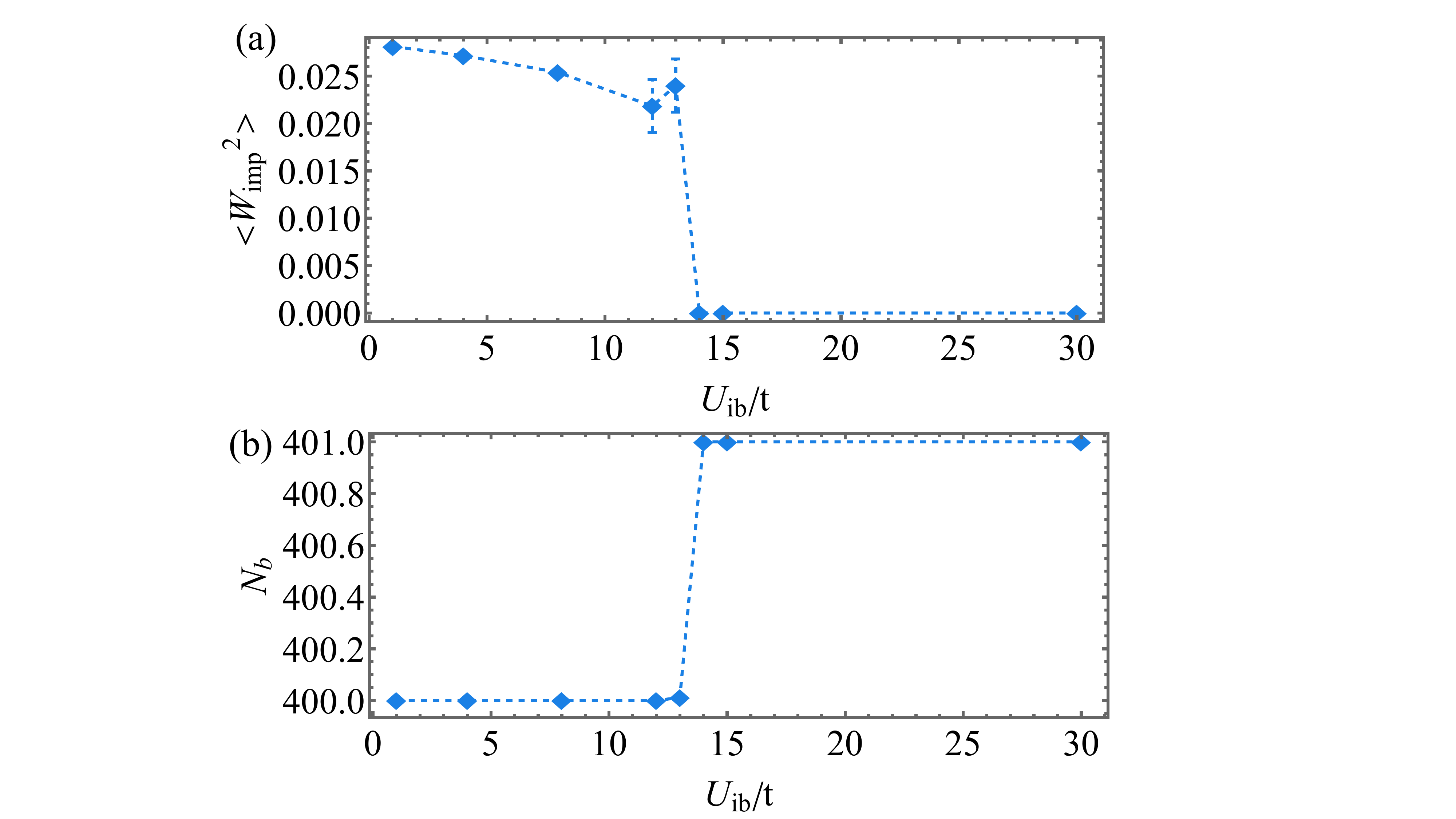}
    \caption{\textbf{Impurity coherence and bath particle number
    for attractive impurity-bath coupling in the MI.}
    (a)~Impurity winding number
    $\langle W_{\mathrm{imp}}^2\rangle$,
    showing finite mobility at small $|U_{\mathrm{ib}}|/t$
    (nearly-free defect) and a sharp suppression
    once a particle-defect forms.
    (b)~Total bath particle number
    $N_{\mathrm{b}}$.
    The discrete jump by one particle near
    $|U_{\mathrm{ib}}|/t\!\approx\!13.0$–14.0
    confirms the quantized recruitment of a bath boson
    and the formation of a self-trapped composite defect.}
    \label{fig:MIatt3}
\end{figure}

We now examine the opposite limit of an attractive impurity
($U_{\mathrm{ib}}/t<0$) embedded in the same Mott-insulating background
with $U_{\mathrm{b}}/t=20.0$.
Because the bath is incompressible, density redistribution cannot
occur smoothly as in the SF case; instead, the impurity may only
recruit bath bosons in discrete, integer steps.
Figures~\ref{fig:MIatt1}–\ref{fig:MIatt3} summarize the evolution
of the cumulative bath response, radial-averaged impurity-centered correlator, impurity coherence,
and total bath occupation.

\paragraph*{Weak attraction: nearly-free defect with minimal dressing.}
For $|U_{\mathrm{ib}}|/t\lesssim 13.0$, the impurity behaves as a
\emph{nearly-free defect} in the MI background.
The cumulative bath response $\Delta N(R)$
[Fig.~\ref{fig:MIatt1}(a)] displays only a small positive bump,
with a shallow maximum
$\Delta N_{\max}\!\ll\!0.1$
located at $R(\Delta N_{\max})\approx 4$
[Fig.~\ref{fig:MIatt1}(b)].
This indicates weak, short-range density attraction toward the impurity.
Consistently, the radial-averaged impurity-centered correlator $C_{\mathrm{ib}}(R)$
[Fig.~\ref{fig:MIatt2}]
shows a modest on-site enhancement with negligible tail,
and the on-site contrast $C_{\mathrm{ib}}(0)$ (inset)
remains close to zero. The impurity winding number
$\langle W_{\mathrm{imp}}^2\rangle$
remains finite and only slowly decreases with increasing $|U_{\mathrm{ib}}|/t$
[Fig.~\ref{fig:MIatt3}(a)],
confirming that the impurity moves coherently through the rigid MI
background.
The total bath particle number $N_{\mathrm{b}}$
is pinned at its integer value
[Fig.~\ref{fig:MIatt3}(b)],
establishing that \emph{no} bath particle has yet been self-trapped.
Overall, the impurity forms only a weakly dressed defect,
in sharp contrast to the extended droplets expected in a
compressible superfluid environment.

\paragraph*{Strong attraction: formation of a self-trapped particle defect.}
A qualitative change occurs around
$|U_{\mathrm{ib}}|/t\simeq 13.0$–$14.0$.
The cumulative bath response develops a rapidly growing positive maximum
$\Delta N_{\max}\sim 1$
[Fig.~\ref{fig:MIatt1}(c)],
while its position $R(\Delta N_{\max})$
remains confined to the first a few shells,
$R\approx 1$–$3$
[Fig.~\ref{fig:MIatt1}(b)].
This indicates that the impurity abruptly captures \emph{one} additional bath boson,
i.e., a quantized $+1$ change in the bath occupation, forming a compact particle-like defect.
The clustering is evident in the local correlator:
$C_{\mathrm{ib}}(0)$ increases sharply and reaches a large positive value
[inset of Fig.~\ref{fig:MIatt2}],
while the shape of $C_{\mathrm{ib}}(R)$ shows a strong
on-site accumulation followed by a tiny tail.
Beyond this threshold $|U_{\rm ib}|/t >14,0$, further increasing $|U_{\mathrm{ib}}|/t$
does not enlarge the accumulation cloud, only deepens the onsite $C_{\rm in}(0)$.

\paragraph*{Collapse of coherence and discrete particle recruitment.}
The formation of the particle-defect cluster is accompanied by
a suppression of coherent impurity motion:
$\langle W_{\mathrm{imp}}^2\rangle$ drops to zero
for $|U_{\mathrm{ib}}|/t\gtrsim 14.0$ [Fig.~\ref{fig:MIatt3}(a)].
Meanwhile, the total bath occupation
$N_{\mathrm{b}}$
increases by \emph{exactly one} particle
[Fig.~\ref{fig:MIatt3}(b)],
demonstrating that the impurity binds to an additional bath boson
to form a self-trapped composite defect. This mechanism sharply contrasts with the smooth droplet formation
in a superfluid bath.
In the MI, the lack of compressibility forbids extended halos;
self-trapping proceeds through a \emph{quantized recruitment}
of bath particles.
The resulting self-trapped object is thus best viewed as a
particle defect,
distinct from both polarons (SF) and vacancy defect (repulsive MI).

Together, these results establish two characteristic regimes
for attractive impurities in an incompressible bath:
(i)~a weakly dressed nearly-free defect at small $|U_{\mathrm{ib}}|/t$;
and (ii)~a strongly bound, self-trapped particle defect at larger
$|U_{\mathrm{ib}}|/t$, marked by discrete particle capture and
vanishing impurity winding.

\begin{figure*}[t]
    \centering
    \includegraphics[width=0.9\linewidth]{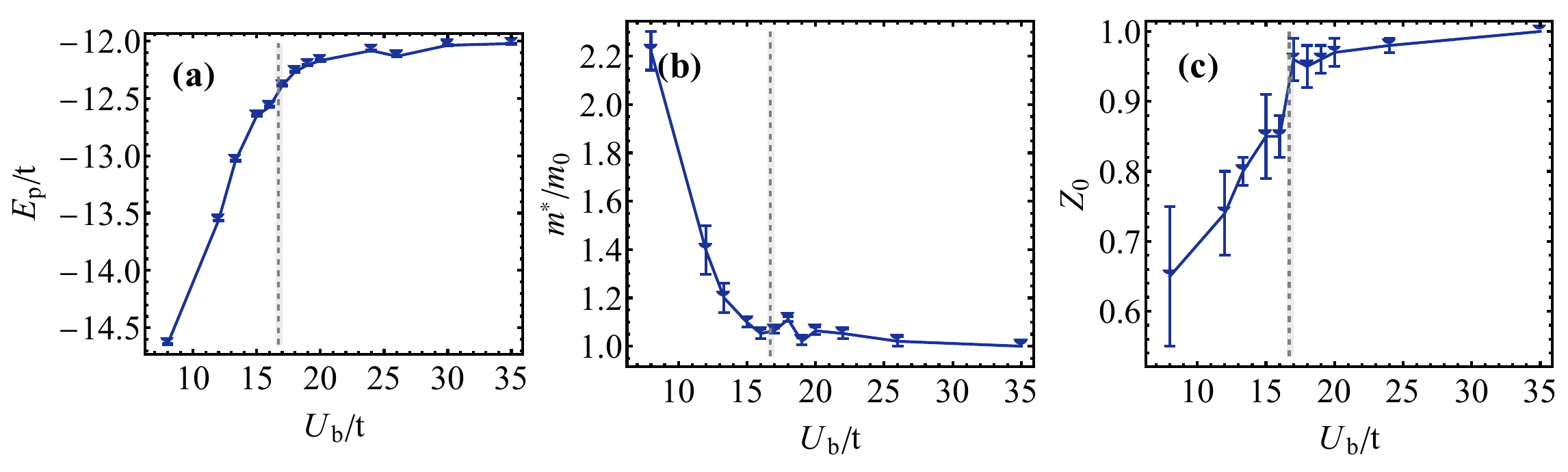}
    \caption{\textbf{Quasiparticle properties of an attractive impurity across the SF--MI transition at fixed impurity-bath coupling $U_{\mathrm{ib}}/t=-8.0$.}
    Ground-state energy $E_{\mathrm{p}}/t$ (a), effective mass $m^*/m_0$
    (b), and quasiparticle weight $Z_0$ (c)
    as functions of bath interaction $U_\mathrm{b}/t$.
    The gray shaded band highlights the SF--MI crossover region
    ($U_\mathrm{b}/t\approx16.5$--$17.0$).
    In the superfluid regime, the impurity behaves as a mobile light
    polaron with finite dressing cloud and moderate mass enhancement.
    As the bath approaches the SF--MI transition, both $m^*/m_0$ and $Z_0$
    approach their bare values, reflecting the compressibility-cotrolled
    collapse of dressing.
    In the deep Mott-insulating regime, the impurity becomes a \emph{nearly-free
    defect}, without forming a self-trapped bound defect or recruiting an
    extra bath boson.
    Unlike the repulsive case at $U_{\mathrm{ib}}/t=8.0$, no quantized defect or vacancy/polaron
    complex is formed.}
    \label{fig:att_vertical}
\end{figure*}

\paragraph*{Contrast with the superfluid bath.}
In the compressible superfluid bath studied in Section~\ref{sec:SF}
the same impurity builds up a \emph{continuous} accumulation halo as
$U_{\mathrm{ib}}/t\!\to\!-\infty$, with no quantized steps in the particle number.
In the Mott-insulating bath considered here, by contrast, incompressibility
prohibits extended dressing and enforces \emph{discrete} recruitment of bath
bosons once binding thresholds are crossed, leading to an integer increase in
the local occupation and a concomitant collapse of coherent impurity motion.

%====================  Section III: Crossing the SF–MI Transition  ====================%
\section{Crossing the superfluid–Mott insulator transition at fixed impurity–bath coupling}
\label{sec:crossing}

\begin{figure}[t]
    \centering
   \includegraphics[width=\linewidth, trim = 30pt 0pt 1120pt 20pt, clip]{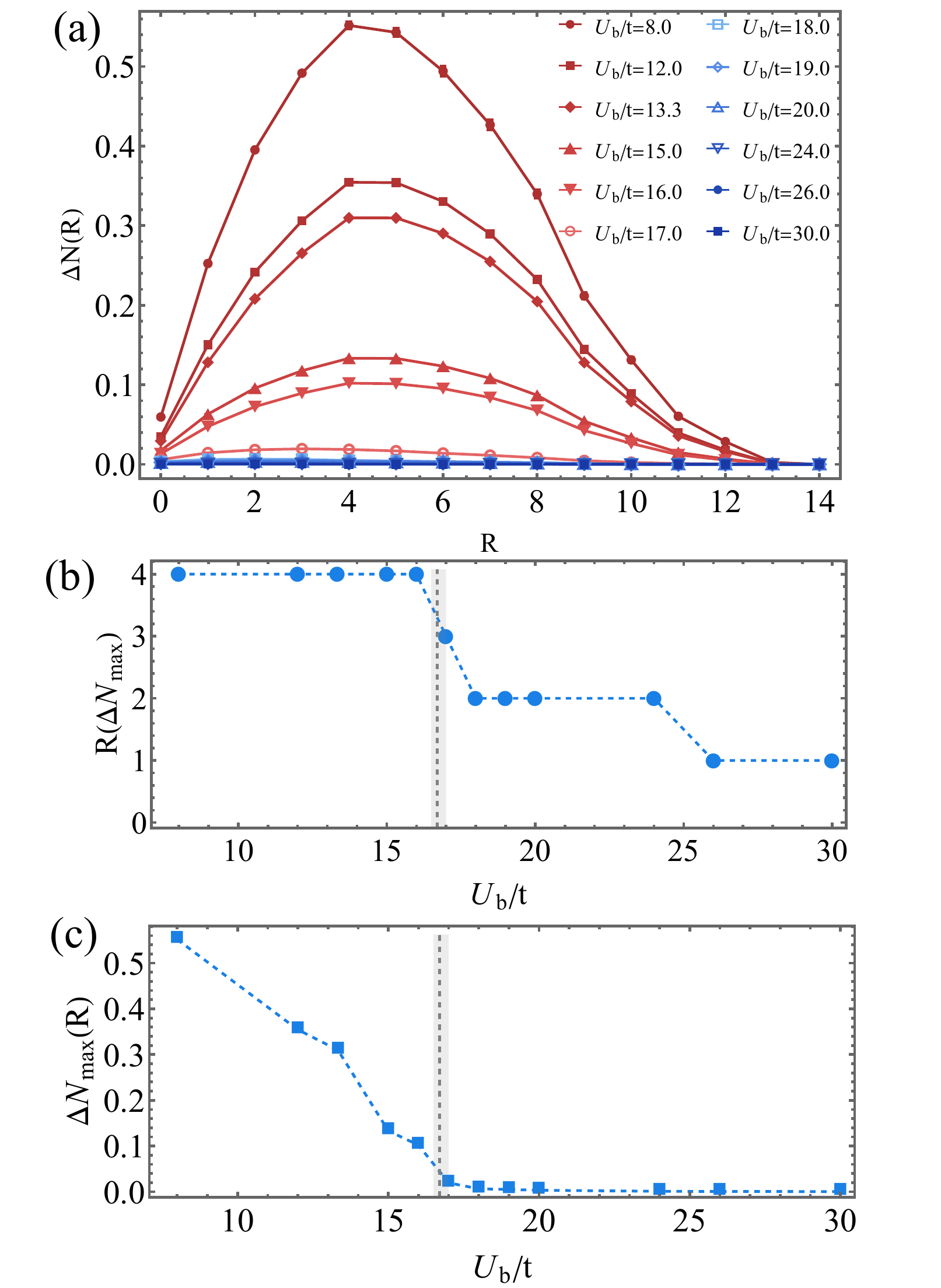}
    \caption{\textbf{Real-space impurity dressing for an attractive impurity across the SF--MI transition at fixed impurity--bath coupling $U_{\mathrm{ib}}/t=-8.0$.}
    (a)~Cumulative bath-density deformation $\Delta N(R)$ for increasing
    $U_{\mathrm{b}}/t$. In the superfluid regime
    ($U_{\mathrm{b}}/t\lesssim 15.0$), the impurity generates an extended
    accumulation halo with maximum $\Delta N_{\mathrm{max}}\approx0.5$
    at $R\approx4$, characteristic of a light polaron with finite spatial dressing.
    As the bath approaches the SF--MI critical region
    ($U_{\mathrm{b}}/t\sim16.7$), both the amplitude and width of the halo
    collapse, reflecting the suppression of bath compressibility and the disappearance
    of long-range screening.
    Deep in the Mott-insulating regime ($U_{\mathrm{b}}/t\gtrsim 18.0$),
    $\Delta N(R)$ becomes almost flat with
    $\Delta N_{\mathrm{max}}\lesssim0.02$ and
    $R(\Delta N_{\mathrm{max}})=1$, indicating a purely local,
    short-range distortion without quantized particle recruitment.
    (b)~Radius of maximum $R(\Delta N_{\mathrm{max}})$ rapidly drops
    from $R\approx4$ in the superfluid to $R=1$ in the Mott-insulating phase,
    confirming that dressing becomes strictly local once compressibility vanishes.
    (c)~Magnitude of maximum $\Delta N_{\mathrm{max}}$
    with increasing $U_{\mathrm{b}}/t$.
    Its smooth decay and nearly zero value demonstrate that the impurity
    remains a weakly dressed, nearly free defect, never reaching the
    particle-defect regime for this impurity--bath coupling strength $U_{\mathrm{ib}}/t=-8.0$.}
    \label{fig:att_vertical1}
\end{figure}

We now take a complementary viewpoint to the interaction-driven paths discussed
above: instead of fixing the bath interaction $U_{\mathrm{b}}/t$ and varying impurity--bath coupling $U_{\mathrm{ib}}/t$, we
perform ``vertical cuts'' in the $(U_{\mathrm{b}}/t,U_{\mathrm{ib}}/t)$ plane by
scanning the bath interaction $U_{\mathrm{b}}/t$ while keeping the
impurity--bath coupling $U_{\mathrm{ib}}/t$ fixed. This strategy directly reveals how impurity
properties evolve as the \emph{bath medium} crosses the superfluid--Mott-insulator
transition, and allows us to isolate the role of bath compressibility from that
of the impurity--bath coupling strength itself.

We begin with the representative intermediate attractive impurity--bath coupling
$U_{\mathrm{ib}}/t=-8.0$, which in a superfluid bath lies close
to the self-trapping threshold between the light polaron region and heavy polaron region identified in the phase diagram of our
companion Letter~\cite{chaoletter}. This choice provides an ideal starting point to trace how the impurity’s
coherence, mobility, and dressing cloud collapse as the bath loses
compressibility: as $U_{\mathrm{b}}/t$ is increased from the
deep-superfluid regime ($U_{\mathrm{b}}/t=8.0$–16.0), through the
critical region ($U_{\mathrm{b}}/t\simeq16.7$), and into the
Mott-insulating regime ($U_{\mathrm{b}}/t=17.0$–24.0), the impurity evolves
from a well-defined light polaron in the SF to a nearly free defect in the MI.

\paragraph*{Overview and observables.}
Along this compressibility-driven trajectory at fixed
$U_{\mathrm{ib}}/t=-8.0$, we monitor both quasiparticle-level and
bath-level quantities. On the impurity side we extract (i) the ground-state
energy $E_{\mathrm{p}}$, (ii) the effective mass $m^{*}/m_0$ from the single-impurity Green's
function, (iii) the quasiparticle residue $Z_0$, and (iv) the impurity
winding-number estimator $\langle W_{\mathrm{imp}}^{2}\rangle$. On the bath side we measure (i) the superfluid density
$\rho_{\mathrm{b}}$ from global winding statistics, which tracks macroscopic
phase coherence and stiffness, and (ii) the radial-averaged impurity-centered density response
encoded in $C_{\mathrm{ib}}(R)$ and its cumulative bath-density deformation
$\Delta N(R)$, which reveal the range and amplitude of the
dressing cloud. Taken together, these observables provide a unified microscopic
picture linking global coherence, local density structure, and quasiparticle
properties across the compressibility-controlled SF--MI transition.

In contrast to the repulsive case analyzed in the companion Letter~\cite{chaoletter}, where a
moderate repulsion $U_{\mathrm{ib}}/t=+8.0$ drives the impurity
into a self-trapped defect in a deep Mott-insulating background, we now examine the attractive case at fixed impurity--bath coupling
$U_{\mathrm{ib}}/t = -8.0$, and vary the bath interaction $U_\mathrm{b}/t$
across the superfluid--Mott-insulator transition.
This vertical trajectory is qualitatively different from the repulsive one:
because the attraction is moderate and the bath becomes increasingly
incompressible as $U_\mathrm{b}/t$ increases, the impurity can no longer
sustain a heavy polaron nor form a bound cluster with multiple bath
bosons.
Instead, its dressing cloud continuously shrinks, and ultimately the
impurity crosses over into a nearly free defect even in the deep Mott-insulating regime.
Importantly, at this coupling strength the impurity \emph{never binds into
a particle defect} and \emph{never recruits an additional bath particle};
it remains mobile, with only minimal local density distortion.

\paragraph*{Quasiparticle properties across the SF--MI transition.}
Figure~\ref{fig:att_vertical} shows the evolution of the ground-state
energy $E_{\mathrm{p}}/t$, effective mass $m^*/m_0$, and quasiparticle residue $Z_0$
as functions of $U_\mathrm{b}/t$.
In the superfluid regime ($U_\mathrm{b}/t \lesssim16.0$),
the impurity behaves as a light polaron characterized by extended
density accumulation, moderately enhanced mass $m^*/m_0\simeq1.1$--$2.5$,
and quasiparticle weight $Z_0\simeq0.7$--$0.9$.
As the bath becomes less compressible and approaches the critical region
($U_\mathrm{b}/t\approx16.7$, shaded band),
both $m^*/m_0$ and $Z_0$ rapidly approach their bare values,
reflecting a compressibility-controlled collapse of polaronic dressing. Deep in the Mott-insulating regime ($U_\mathrm{b}/t\geq18.0$),
bath density fluctuations are frozen, and the impurity behaves as a
nearly free defect with $m^*/m_0\approx1$, $Z_0\to1$, and only a small,
strictly local density enhancement.

\paragraph*{Bath responses across the SF--MI transition.}
Figure~\ref{fig:att_vertical1}(a) shows the impurity-centered cumulative bath-density deformation
$\Delta N(R)$ for a sequence of increasing $U_{\mathrm{b}}/t$ at
fixed $U_{\mathrm{ib}}/t=-8.0$.
In the superfluid regime $U_{\mathrm{b}}/t\lesssim16.0$, the
impurity induces a smooth, spatially extended density \emph{accumulation halo},
with a broad maximum $\Delta N_{\mathrm{max}}\approx0.5$ located at
$R\approx4$.
This reflects a light polaron with coherent mobility and an extended dressing
cloud.
As $U_{\mathrm{b}}/t$ approaches the SF--MI critical region
($U_{\mathrm{b}}/t\approx16.7$), both the amplitude
$\Delta N_{\mathrm{max}}$ and the radius $R(\Delta N_{\mathrm{max}})$ shrink
rapidly [Figs.~\ref{fig:att_vertical1}(b,c)], signaling a
\emph{compressibility-controlled collapse} of extended dressing:
the halo can no longer spread through an increasingly rigid bath.

Deep in the Mott-insulating regime ($U_{\mathrm{b}}/t\gtrsim18.0$),
$\Delta N(R)$ becomes almost flat, with
$\Delta N_{\mathrm{max}}\lesssim0.02$ and
$R(\Delta N_{\mathrm{max}})=1$ or 2, indicating that the impurity induces only
a very weak, short-range density enhancement confined to its vicinity.
The tiny amplitude confirms the absence of quantized particle recruitment:
no additional bosons are bound to the impurity, and thus no bound cluster or
particle defect forms.
Instead, the impurity remains a \emph{nearly free defect} propagating in an
essentially rigid background.
Thus for $U_{\mathrm{ib}}/t=-8.0$ the impurity never crosses into a
particle-defect regime, even deep inside the Mott-insulating phase.

These observations provide an important contrast to the repulsive case at
similar impurity--bath coupling $U_{\mathrm{ib}}/t =8.0$ in the MI, where a vacancy defect emerges once the bath becomes
incompressible.
For attractive $U_{\mathrm{ib}}/t=-8.0$, the impurity is energetically unable to
attract and bind a full additional boson because of the large interaction gap
of the Mott insulator; the local distortion remains bounded by
$\Delta N_{\mathrm{max}}\lesssim 0.02$, and no quantized particle capture or cluster
formation occurs.

\begin{figure}[t]
    \centering
\includegraphics[width=\linewidth, trim = 200pt 5pt 300pt 78pt, clip]{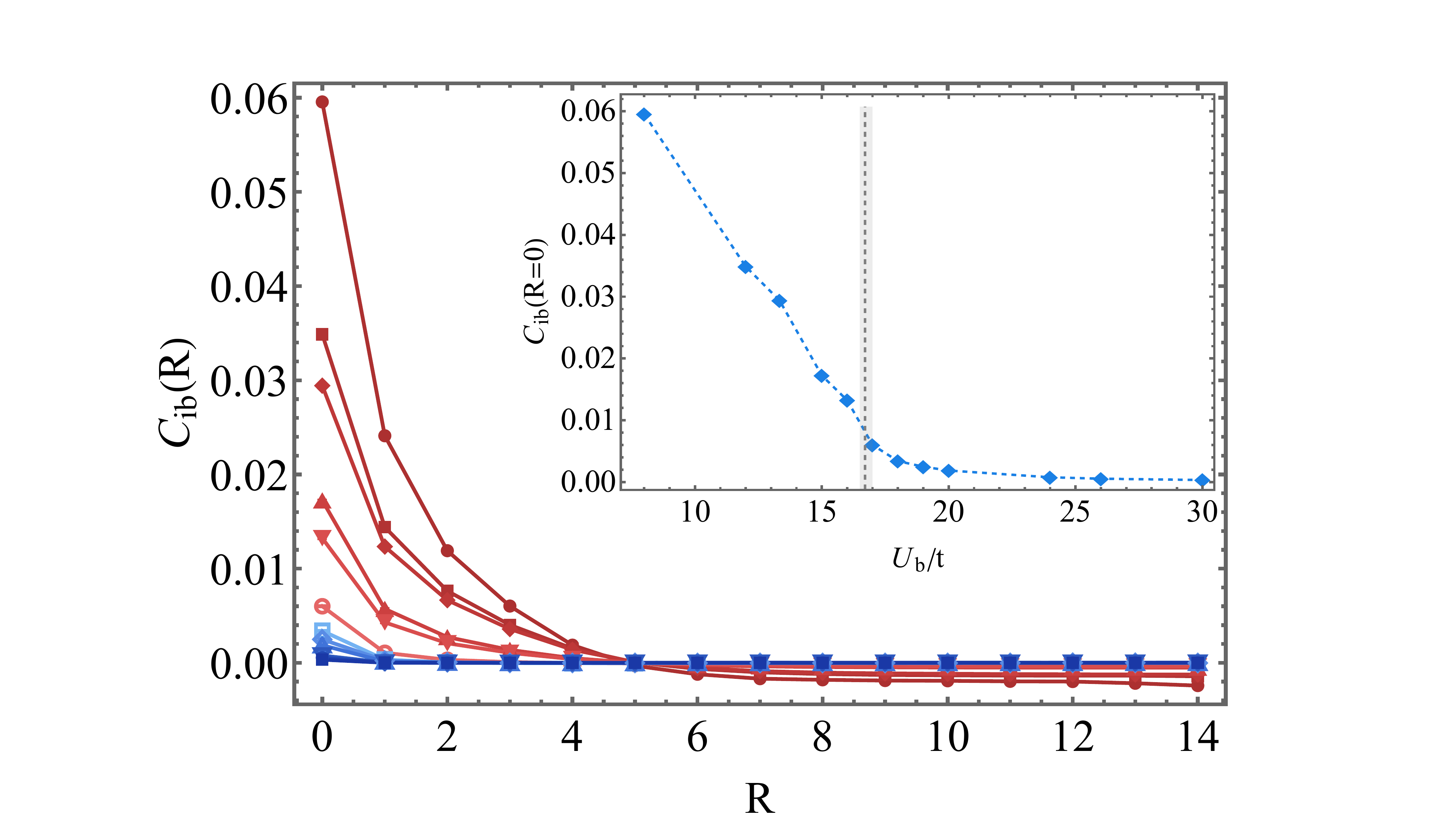}
    \caption{\textbf{Radial-averaged impurity-centered correlator $C_{\mathrm{ib}}(R)$ at fixed attractive coupling
    $U_{\mathrm{ib}}/t=-8.0$ across the SF--MI transition.}
    Main panel: Radial impurity-centered correlator
    $C_{\mathrm{ib}}(R)$ for different bath interactions.
    In the superfluid regime ($U_{\mathrm{b}}/t=8.0$--12.0),
    $C_{\mathrm{ib}}(R)$ shows a smooth, extended positive halo,
    characteristic of a mobile light polaron.
    As $U_{\mathrm{b}}/t$ increases, both the amplitude and range of
    $C_{\mathrm{ib}}(R)$ are continuously suppressed, demonstrating
    the compressibility-driven collapse of bath dressing.
    Deep in the Mott-insulating regime ($U_{\mathrm{b}}/t\ge 22.0$), the correlator
    becomes strictly short-ranged and nearly vanishes,
    indicating a \emph{nearly free defect}
    with no particle-defect formation.
    Inset: On-site contrast $C_{\mathrm{ib}}(0)$ as a function of
    $U_{\mathrm{b}}/t$.
    Its monotonic decay to nearly zero confirms that, at
    moderate attraction ($U_{\mathrm{ib}}/t=-8.0$),
    the impurity never recruits one additional bath boson and
    does not form a self-trapped particle defect.
    }
    \label{fig:att_cib}
\end{figure}

Figure~\ref{fig:att_cib} further characterizes the radial-centered impurity--bath correlator $C_{\mathrm{ib}}(R)$ along the vertical cut at fixed $U_{\mathrm{ib}}/t=-8.0$.
The radial-averaged correlator $C_{\mathrm{ib}}(R)$ in the main panel shows a pronounced
on-site accumulation in the weakly interacting superfluid
($U_{\mathrm{b}}/t= 8.0$), followed by a broad positive tail,
consistent with the extended density halo seen in $\Delta N(R)$ and indicative
of a light polaron with substantial bath dressing.
As $U_{\mathrm{b}}/t$ increases toward the SF--MI boundary, both the amplitude and
the range of $C_{\mathrm{ib}}(R)$ are continuously suppressed, reflecting the
compressibility-controlled collapse of the dressing cloud: an increasingly
rigid bath can no longer support long-ranged density redistribution.

Once the bath enters the Mott-insulating regime ($U_{\mathrm{b}}/t\gtrsim 17.0$),
$C_{\mathrm{ib}}(R)$ becomes extremely short-ranged and decays to zero within
one lattice spacing.
The inset highlights the on-site contrast $C_{\mathrm{ib}}(0)$, which decreases
monotonically with $U_{\mathrm{b}}/t$ and reaches
$|C_{\mathrm{ib}}(0)|\lesssim 10^{-3}$ for $U_{\mathrm{b}}/t>22.0$.
This tiny local signal confirms that, for $U_{\mathrm{ib}}/t=-8.0$, the impurity
does not bind any additional bath bosons in the Mott-insulating phase: no quantized
particle defect is formed, and the impurity remains a
weakly dressed, nearly free defect embedded in an incompressible background.
This behavior stands in sharp contrast to the strong-attraction regime
($|U_{\mathrm{ib}}|/t\gtrsim 12.0$) discussed in
Sec.~\ref{subsec:MI_attractive}, where a quantized particle defect with
$\Delta N_{\max}\simeq 1$ is stabilized in the MI phase.

\begin{figure}[t]
    \centering
\includegraphics[width=\linewidth, trim = 200pt 10pt 360pt 40pt, clip]{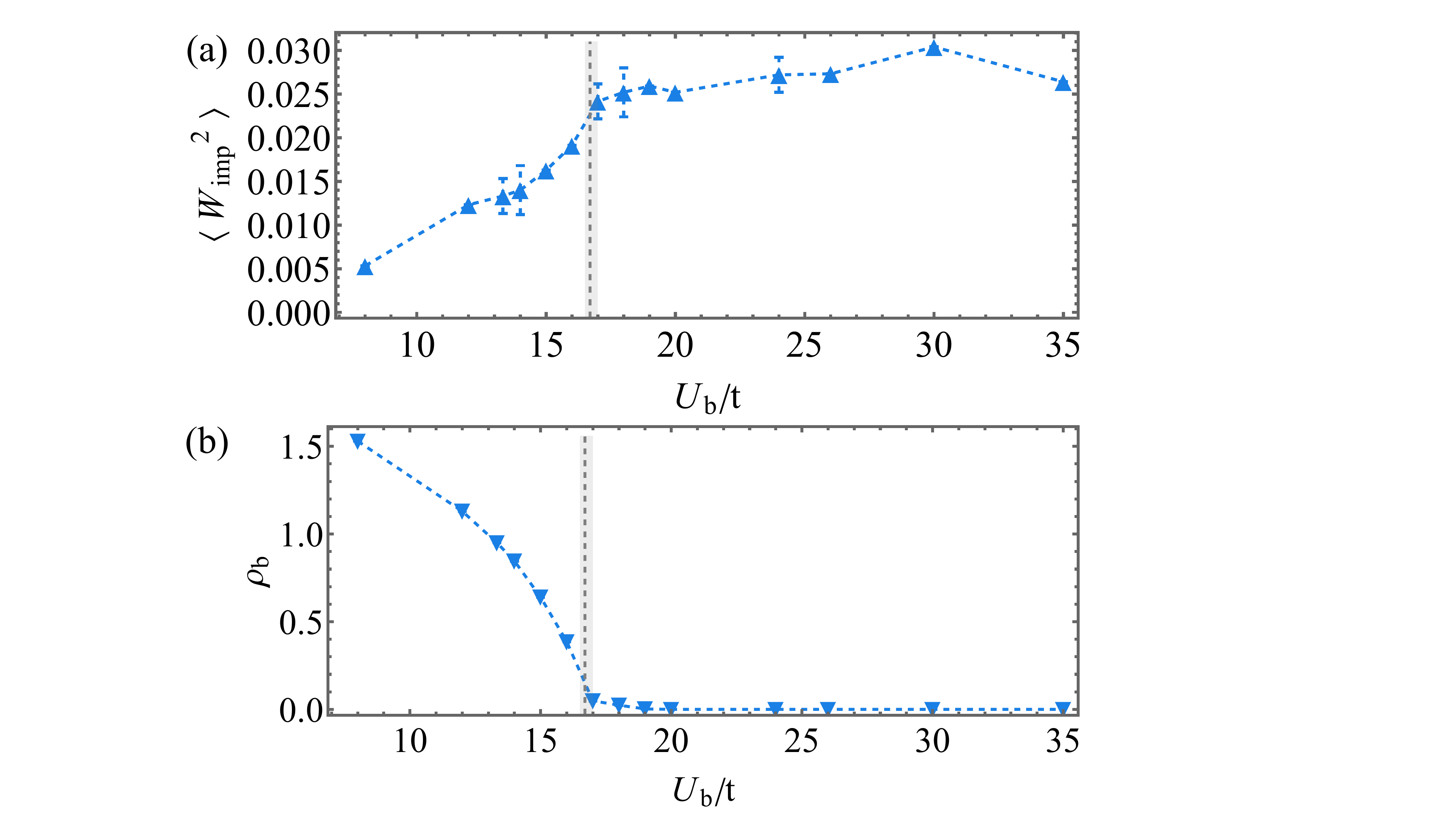}
    \caption{\textbf{Compressibility-controlled crossover of an attractive impurity across the SF--MI transition at fixed $U_{\mathrm{ib}}/t=-8.0$.}
    (a) Impurity winding-number estimator
    $\langle W_{\mathrm{imp}}^{2}\rangle$ as a function of
    bath interaction $U_{\mathrm{b}}/t$.
    In the superfluid regime ($U_{\mathrm{b}}/t<16.7$),
    $\langle W_{\mathrm{imp}}^{2}\rangle$ is finite, indicating a mobile
    light polaron.
    Across the SF--MI transition the winding \emph{increases} and remains finite even
    deep in the Mott-insulating regime, showing that the impurity crosses over to a nearly
    free defect rather than self-trapping into a particle defect.
    (b) Bath superfluid density $\rho_{\mathrm{b}}$ extracted from global
    winding statistics, demonstrating the loss of phase stiffness across the
    superfluid--Mott-insulator transition.
    The collapse of $\rho_{\mathrm{b}}$ coincides with the disappearance of extended dressing,
    while the finite $\langle W_{\mathrm{imp}}^{2}\rangle$ in panel~(a) highlights the
    persistence of coherent impurity motion in the Mott-insulating regime.
    }
    \label{fig:attWinding}
\end{figure}

The winding estimator of the impurity and bath superfluid density in
Fig.~\ref{fig:attWinding} provide a complementary view of the same
compressibility-controlled crossover.
In the superfluid regime the impurity winding
$\langle W_{\mathrm{imp}}^{2}\rangle$ is clearly finite, consistent with a
mobile light polaron dressed by an extended density halo.
As $U_{\mathrm{b}}/t$ is increased and the bath superfluid density
$\rho_{\mathrm{b}}$ collapses at the SF--MI transition, the impurity winding
increases and remains finite for the largest $U_{\mathrm{b}}/t$ considered.
This confirms that, for $U_{\mathrm{ib}}/t=-8.0$, the impurity evolves into a
nearly free defect whose weak residual dressing is captured by the small
cumulative bath-density deformation in $\Delta N(R)$ and the short-ranged correlator
$C_{\mathrm{ib}}(R)$.

\section{Summary of main results}
Taken together, our results provide a unified microscopic picture of how a
single impurity self-traps in the two-dimensional Bose--Hubbard model when the
impurity--bath coupling is attractive. Along the interaction-driven trajectory
in a fixed superfluid bath ($U_{\mathrm{b}}/t=13.3$), increasing
$|U_{\mathrm{ib}}|/t$ drives a continuous winding-collapse crossover from a
mobile light polaron with an extended dressing cloud, to a compact heavy
polaron, and finally to a tightly bound multi-particle cluster. %This evolution is captured consistently by the winding number estimator of the impurity, the real-space impurity-centered correlators $C_{\mathrm{ib}}(R)$ and cumulative bath-density deformation $\Delta N(R)$, and the quasiparticle parameters $E_{\mathrm{p}}$, $m^{*}/m_0$, and $Z_0$: the impurity winding is suppressed, the cloud contracts, the effective mass grows, and the residue is strongly reduced while the bath remains globally superfluid.

In the Mott-insulating regime, incompressibility qualitatively reshapes the
impurity physics. For sufficiently strong repulsive and attractive couplings,
the impurity binds into quantized vacancy and particle defects,
respectively: the impurity winding collapses, $\Delta N(R)$ saturates at
integer values (e.g., $\Delta N_{\min}\simeq -1$ for a vacancy and
$\Delta N_{\max}\simeq 1$ for a particle defect), and the deformation becomes
confined to within one or two lattice spacings. The impurity thus turns into a
fully self-trapped defect, and the Mott-insulating bath accommodates it by carrying an
integer particle deficit or surplus in its vicinity, directly reflecting the
incompressibility of the Mott-insulating background.

A complementary viewpoint is provided by compressibility-controlled trajectories at
fixed impurity--bath coupling, exemplified here by the intermediate attraction
$U_{\mathrm{ib}}/t=-8.0$. As the bath interaction
$U_{\mathrm{b}}/t$ is increased from the deep-superfluid regime,
through the critical region, and into the deep Mott-insulating regime, the bath superfluid
stiffness collapses and the extended density halo around the
impurity shrinks and eventually disappears. At the same time, the impurity
winding increases and the deformation amplitude
$\Delta N_{\max}$ decays below $0.02$, with no quantized particle capture.
Along such trajectories, the impurity dynamics are governed by a purely compressibility-controlled suppression of extended dressing, and the
impurity crosses over from a light polaron in a compressible superfluid bath
to a weakly dressed, nearly free defect in an incompressible Mott insulator.
The comparison of these regimes highlights that impurity self-trapping in the
Bose--Hubbard model can proceed either via interaction-driven winding collapse
in a superfluid bath, or via a compressibility-controlled collapse of
polaronic dressing when the bath is tuned across the SF--MI transition.

\section{Experimental realization}
\label{sec:experiment}

The two-component Bose--Hubbard Hamiltonian in Eq.~\eqref{eq:Hamiltonian}
can be implemented with state-of-the-art ultracold bosonic mixtures in
optical lattices. A natural realization uses two hyperfine states of
$^{87}$Rb or $^{41}$K loaded into a square optical lattice, where one
component (the bath) realizes a Bose--Hubbard system with a tunable ratio
$U_{\mathrm{b}}/t_{\mathrm{b}}$ controlled by the lattice depth and
scattering length~\cite{JakschPRL1998,GreinerNature2002,KrutitskyPhysRep2016,GrossBlochScience2017}.
A dilute population of atoms in a second hyperfine state then plays the role
of mobile impurities with tunneling amplitude $t_{\mathrm{imp}}$ and an
on-site interaction $U_{\mathrm{ib}}$ with the bath species.

The impurity--bath coupling $U_{\mathrm{ib}}$ can be tuned in both magnitude
and sign using interspecies Feshbach resonances in bosonic mixtures,
providing access to the repulsive and attractive regimes discussed in this
work~\cite{BestPRL2009}. Independent control of the impurity hopping
$t_{\mathrm{imp}}$ is achievable with species-selective or
state-dependent optical lattices, which engineer different lattice depths
for the two components and thus realize tunable ratios
$t_{\mathrm{imp}}/t_{\mathrm{b}}$~\cite{LeBlancPRA2007,MandelPRL2003}. Within these established techniques, the parameter ranges explored in our
QMC simulations---covering superfluid and Mott-insulating baths and
intermediate impurity--bath couplings---are well matched to those already
realized in current experiments on lattice bosons.

Single-site--resolved imaging with quantum gas microscopes provides direct
access to the real-space observables introduced in this work~\cite{BakrScience2010,ShersonNature2010,FukuharaNatPhys2013}.
In particular, the impurity-centered correlator $C_{\mathrm{ib}}(\mathbf r)$
can be reconstructed from conditional density maps by post-selecting
snapshots in which the impurity is found at position
$\mathbf r_{\mathrm{imp}}$ and averaging the bath occupation at
$\mathbf r_{\mathrm{imp}}+\mathbf r$. Radial integration of these maps
yields the cumulative deformation $\Delta N(R)$, directly visualizing the
formation of depletion bubbles or bound clusters around the impurity.
Moreover, repeated preparation and readout of the impurity trajectory
would allow one to access winding-number statistics and thereby probe the
crossover from coherent motion to self-trapping into defect-bound states.

Taken together, the combination of tunable lattice depth, controllable
interspecies interactions, and single-site readout provides a realistic
experimental route to explore all impurity regimes identified in our QMC
study: from mobile light polarons through heavy polarons to saturated
bubbles (for repulsive impurity--bath couplings) or bound clusters (for
attractive couplings) in the superfluid bath, and from nearly free defects
to fully self-trapped vacancy- or particle-like defects in the Mott-insulating
regime.

\section{Discussion and Outlook}
\label{outlook}

The present work provides a unified microscopic picture of impurity self-trapping
across the compressible superfluid phase and the incompressible Mott-insulating phase within the two-dimensional
Bose--Hubbard model.
By combining worldline winding diagnostics with real-space bath deformations,
we have established that impurity self-trapping is controlled by two distinct mechanisms
depending on the bath compressibility and impurity--bath coupling.

In the superfluid regime, where long-wavelength phase fluctuations are gapless
and the bath remains fully compressible, the buildup of dressing is a
continuous process.
As the impurity--bath coupling $|U_{\mathrm{ib}}|/t$ is increased at fixed $U_{\mathrm{b}}/t$,
the impurity’s winding number and the associated coherent motion are gradually
suppressed: the winding estimator of the impurity decreases and eventually collapses as the impurity enters a self-trapped regime.
In parallel, the cumulative bath-density deformation $\Delta N(R)$ evolves smoothly from a weak,
extended halo, reflected by a broad, low-amplitude $C_{\mathrm{ib}}(R)$, to a compact, high-contrast structure in which the
response $C_{\mathrm{ib}}(R)$ is concentrated within one or two lattice spacings.
On the repulsive side, this limiting configuration corresponds to a saturated
depletion bubble with $\Delta N_{\min}\simeq -1$, whereas on the attractive
side it becomes a dense bound cluster with $\Delta N_{\max}\gtrsim 1$.
Throughout this evolution the bath remains globally superfluid, so the
self-trapping is controlled entirely by the interaction-driven collapse of
winding and dressing, rather than by any loss of phase coherence in the bath.

In contrast, deep in the Mott-insulating regime, the situation changes qualitatively.
The bath becomes incompressible, long-wavelength fluctuations are frozen,
and only local particle--hole excitations with discrete energies remain.
The impurity can no longer dress itself continuously;
instead, self-trapping proceeds through defect binding with
quantized density rearrangements:
for repulsive coupling, a single bath boson is expelled from the impurity vicinity,
producing a vacancy defect with $\Delta N_{\min}\!\approx\!-1$;
for attractive coupling, one boson is captured to form a
particle defect with $\Delta N_{\max}\!\approx\!1$.
These two regimes represent mirror images of each other,
illustrating how the sign of impurity--bath coupling $U_{\mathrm{ib}}/t$ determines whether
the impurity locally depletes or enriches the background.
The collapse of $\langle W_{\mathrm{imp}}^2\rangle$,
together with the discrete change of the bath occupation,
demonstrates that the impurity’s coherence vanishes abruptly
and the underlying density response is quantized.

From a broader perspective, our results suggest that a single impurity
acts as a local quantum probe of the many-body gap.
The correlations between the impurity winding, the extremal cumulative
deformation $\Delta N_{\rm ext}$ (either $\Delta N_{\min}$ or
$\Delta N_{\max}$), and the global bath compressibility provide a
sensitive diagnostic of the onset of incompressibility:
as the gap opens, extended dressing collapses and the integrated density
deformation becomes quantized.
This type of impurity-based diagnosis is conceptually similar in spirit
to local spectroscopy and impurity thermometry in quantum gases, but is
realized here entirely within an equilibrium path-integral framework.
The same observables could be measured directly using
site-resolved quantum gas microscopy:
$C_{\mathrm{ib}}(\mathbf r)$ can be extracted from conditional density maps
averaged over snapshots in which the impurity occupies site
$\mathbf r_{\mathrm{imp}}$, while the cumulative bath-density response
$\Delta N(R)$ can be reconstructed from the radially averaged
impurity-centered correlator $C_{\mathrm{ib}}(\mathbf r)$.
Tracking the impurity trajectory over repeated experimental runs
would give access to winding-number statistics and hence to the
crossover from mobile to self-trapped regimes.

Several extensions of this work naturally arise.
First, finite-temperature effects may reveal the thermal activation
of particle--hole excitations in the Mott-insulating background,
leading to a temperature-dependent de-self-trapping crossover.
With multiple impurities, one can probe bath-mediated impurity–impurity interactions: in the superfluid they emerge as weak, long-range effective forces transmitted by compressible fluctuations, whereas in the Mott insulator they can manifest as tightly bound composite states formed via defect binding.
Third, relaxing the constraint $t_{\mathrm{a}}\!=\!t_{\mathrm{b}}$
introduces an additional tuning knob:
in the limit $t_{\mathrm{a}}\!\ll\!t_{\mathrm{b}}$
the impurity becomes effectively heavy,
approaching the static defect problem,
whereas $t_{\mathrm{a}}\!\gg\!t_{\mathrm{b}}$ realizes a fast impurity
moving through a slow, strongly correlated bath.
Finally, by focusing on the two-impurity sector one can address
bipolaron formation directly.
Even in the simplest case without explicit impurity--impurity
interactions, bath-mediated attraction may bind two impurities into a
composite object whose binding energy and effective mass can be mapped
across the $(U_{\mathrm{ib}}/t, U_{\mathrm{b}}/t)$ plane.
Including a tunable direct impurity--impurity coupling would further
allow one to disentangle bath-mediated interactions and bare impurity-impurity interactions, and remains
fully accessible within the same sign-problem-free QMC framework.

\noindent\textbf{Acknowledgments--}
We are grateful to Guido Pupillo for bringing this idea to our attention and for helpful discussions. We thank Nikolay Prokof'ev for insightful discussions.

\bibliography{impurity}

\end{document}